\documentclass[%
footinbib,
superscriptaddress,
twocolumn,
amsmath,amssymb,
aps,
]{revtex4-1}

\usepackage{graphicx}
\usepackage{dcolumn}
\usepackage{bm}
\usepackage[justification=RaggedRight]{caption}
\usepackage{subcaption}
\usepackage[utf8]{inputenc}
\usepackage[T1]{fontenc}
\usepackage{color,soul} 

\usepackage{xcolor}		
\usepackage[english]{babel} 

\newbox{\bigpicturebox}
\DeclareMathOperator\erf{erf}
\DeclareMathOperator\erfc{erfc}

\begin{document}

\preprint{APS/123-QED}

\title{Interactions and migration rescuing ecological diversity 
}

\author{Giulia Garcia Lorenzana}
\affiliation{Laboratoire de Physique de l'\'Ecole normale sup\'erieure, ENS, Universit\'e PSL, CNRS, Sorbonne Universit\'e, Universit\'e de Paris F-75005 Paris, France}
\affiliation{Laboratoire Matière et Systèmes Complexes (MSC), Université Paris Cité, CNRS, 75013 Paris, France}

\author{Ada Altieri}
\affiliation{Laboratoire Matière et Systèmes Complexes (MSC), Université Paris Cité, CNRS, 75013 Paris, France}
\author{Giulio Biroli}
\affiliation{Laboratoire de Physique de l'\'Ecole normale sup\'erieure, ENS, Universit\'e PSL, CNRS, Sorbonne Universit\'e, Universit\'e de Paris F-75005 Paris, France}

\begin{abstract}


How diversity is maintained in natural ecosystems is a long-standing question in Theoretical Ecology. By studying a system that combines ecological dynamics, heterogeneous interactions and spatial structure, we uncover a new mechanism for the survival of diversity-rich ecosystems in the presence of demographic fluctuations. For a single species, one finds a continuous phase transition between an extinction and a survival state, that falls into the universality class of Directed Percolation. Here we show that the case of many species with heterogeneous interactions is different and richer. By merging theory and simulations, we demonstrate that with sufficiently strong demographic noise, the system exhibits behavior akin to the single-species case, undergoing a continuous transition. Conversely, at low demographic noise, we observe unique features indicative of the ecosystem's complexity. The combined effects of the heterogeneity in the interaction network and migration enable the community to thrive, even in situations where demographic noise would lead to the extinction of isolated species. The emergence of mutualism induces the development of global bistability, accompanied by sudden tipping points. We present a way to predict the catastrophic shift from high diversity to extinction by probing responses to perturbations as an early warning signal.

\end{abstract}

\maketitle

\section{Introduction} 

Community ecology explores how the interactions between different species shape the diversity-rich ecosystems that characterize the natural world. Understanding the main mechanisms at play is a challenge that spans different scientific fields and it is relevant for human health \cite{lozupone2012}. 

There are three salient facts that one has to take into account in this endeavor. Many ecosystems of interest are {\it species-rich}. The interactions between these large sets of species, and the induced ecological dynamics, can lead to complex dynamical behaviors such as chaos and a very large number of possible equilibria \cite{Kessler2015, bunin2017, biroli2018, altieri2021, galla2018, rogers2022, gross2005}. Many ecosystems are {\it spatially extended}: the ecological dynamics takes place at some local scale, but individuals can then explore different spatial locations through migration \cite{leibold2004}. 
This can lead to the appearance of complex ecological phenomena, such as traveling activity fronts, pattern formation, and persistent chaotic dynamics \cite{hassell1991, mobilia2007, olmeda2023, dobramysl2018, Roy2020, pearce2020, denk2022, baron2020, leemput2015}. Ecosystems are subject to {\it noise}, in particular environmental and demographic (due to stochasticity in births and deaths). Both noises induce fluctuations which are a key factor in determining abundances distributions, and their time-dependence \cite{may1974, grilli2020, azaele2016, kamenev2008, larroya2023, vasseur2004, petchey1997,realpe-gomez2013, bell2000, shoemaker2020, peruzzo2020}.
Understanding the interplay between these three properties of ecosystems is essential for answering many central questions in community ecology. 

In this work, we consider spatially extended species-rich ecosystems subject to demographic noise. 
We will consider populations that are large but spatially structured, so that demographic fluctuations globally average out, but they have an important effect on the local dynamics.
This is for example the case in semi-arid ecosystems: the total number of plants is such that global fluctuations are negligible, but at the local level stochasticity can play a fundamental role \cite{realpe-gomez2013}.
Our aim is to understand how in these cases interactions and spatial migration can allow for large diversity and finite abundances despite the adversarial role of demographic noise. In fact, in an isolated community demographic noise leads to extinctions, irreversibly reducing the ecosystem's diversity until there are no species left \cite{bell2000}.

Previous works, following the classical theory of Island Biogeography by MacArthur and Wilson \cite{macarthur1967}, proposed as a rescuing mechanism the immigration from a static reservoir (or "mainland", when thinking of an island-mainland system)  \cite{bell2000, Kessler2015, biroli2018, hu2022, garcialorenzana2022}. Nevertheless, this approach simply shifts the question from how diversity is maintained on the island to its maintenance on the mainland.
Here we use a different approach. We consider ecosystems as a network of ecological communities (\textit{a metacommunity}) coupled by passive dispersal.  In this case, the immigration rates are not externally imposed, but they are the result of the internal dynamics.
If a species goes locally extinct in one of the communities, immigrants from the neighboring ones can re-invade, providing an "insurance" (or "storage") effect \mbox{\cite{loreau_biodiversity_2003, chesson_general_2000}}. 
This makes the possibility of a global extinction much more unlikely, and it can allow the ecosystem to self-sustain at finite abundances and diversity.
The stabilisation of high-diversity states by spatial structure is a very general phenomenon: it can arise in the presence of spatial heterogeneity of environmental conditions \mbox{\cite{loreau_biodiversity_2003, chesson_general_2000, gravel2016, leibold2004, pettersson2021}} or when abundances in different spatial locations exhibit unsynchronized fluctuations \mbox{\cite{Roy2020, pearce2020, mahadevan_spatiotemporal_2023, denk2022}}.
Providing a theory for this mechanism for species-rich ecosystems subject to demographic noise, and assessing the role of interactions, is the main contribution of this work. 

The situation is well understood in the case of a few species, in which depending on the competition between migration and death-birth rates the system is found to be either in a survival or in a extinct state. A transition separates the two regimes \cite{broadbent1957, janssen1997, mobilia2007, dobramysl2018}.
This phase transition falls in the universality class of Directed Percolation, a second-order out-of-equilibrium transition studied in statistical physics and widely used to describe spreading phenomena, from forest fires to epidemics \cite{hinrichsen2000}.  

In a many-species metacommunity with constant competitive interactions, it was recently shown that a similar second-order phase transition takes place and that it also belongs to the Directed Percolation universality class \cite{denk2022}.
Because the transition is continuous with vanishing abundances, interactions, that are quadratic in the abundances, are subleading at the critical point. In consequence, the main mechanism at play in this case is still the competition between migration and death-birth rates.  
We shall show that the scenario for \textit{heterogeneous interactions} is different and goes beyond the directed percolation paradigm. The transition can become discontinuous. 
The ecosystem can exhibit global bistability and tipping points between drastically different alternative states. Upon small changes in the environmental condition, the system can therefore undergo
catastrophic shifts from a state with large diversity and finite abundances to one in which all species are extinct.  As in many other dynamical systems, from coral reefs to arid ecosystems and from Earth's climate to financial markets \cite{scheffer2001, kefi2007, lenton2008, bouchaud2013}, it is important to find early warning signals of these kinds of transition in order to prevent them. We have identified a specific probe, which is based on the response of the ecosystem to perturbations, and that can be monitored in experiments. Our analytical framework shows that      
interactions play a key role both in the overall scenario and in promoting a self-sustained survival state, in agreement with results obtained for constant mutualistic interactions \cite{denk2023}. Remarkably, in our case, heterogeneous interactions of the pool of species are not necessarily mutualistic on average. It is the ecological dynamics that shapes the ecosystem in a self-sustained phase characterized by emergent mutualistic behavior among the non-extinct species.    

In our work, we make use of several methods developed in statistical physics that are particularly well suited for species-rich ecosystems, which are complex systems formed by many interacting degrees of freedom undergoing stochastic dynamics. To model the heterogeneity in the interactions, we sample the coupling coefficients from a random ensemble. We have thus to deal with "disordered" ecosystems, which can be analyzed by transferring methods from spin-glass theory \cite{mezard1986}.  
This disorder approach, which dates back to May's seminal paper \cite{May1972}, has recently inspired a growing body of work \cite{bunin2017, biroli2018,altieri2021, altieri2022, larroya2023, galla2018, fisher2014, pigani2022, suweis2023} and also received positive experimental confirmations \cite{barbier2018,hu2022}.
Previous works have explored within this framework the effect of heterogeneous interactions \cite{May1972, bunin2017, biroli2018,altieri2021, galla2018}, demographic fluctuations \cite{altieri2021, larroya2023} and spatial structure \cite{denk2022, Roy2020, pearce2020, olmeda2023, baron2020,dobramysl2018}, but the analysis we present here is to our knowledge the first analytical study in which the three ingredients are combined. 

The model we focus on is a disordered Generalized Lotka Volterra (GLV) system of metacommunity subject to demographic noise. For one community, the disordered GLV has been shown to have a rich phase diagram, and to display several dynamical regimes: single equilibrium, multi-stability, and chaos \cite{bunin2017, biroli2018, altieri2021, hu2022, galla2018}. 
We expect this complex behavior also in the case of spatially structured ecosystems \cite{olmeda2023}. In this work, we focus on the moderate-heterogeneity regime in which there is a single stable equilibrium. This allows us to disentangle the multistability due to the fragmentation of the basins of attraction of the ecological dynamics at strong heterogeneity from the bistability of the feedback mechanism between abundance and immigration. Our analysis is performed using a mean-field approximation on the spatial fluctuations, which is equivalent to considering that the community network is a fully connected graph. 

Note that because of their generality, Lotka-Volterra equations have been applied to a variety of fields besides their original ecological interpretation, from immunology to economics and game theory \cite{Behn1992, moran2019, Goodwin1990, Bomze1995}.
Our results could therefore find applications beyond ecology, notably for the study of global bistability and crashes in economy.

\section{The model} 
\label{sec:model}

\begin{figure}[htbp]
\centering
\includegraphics[width=0.4\textwidth]{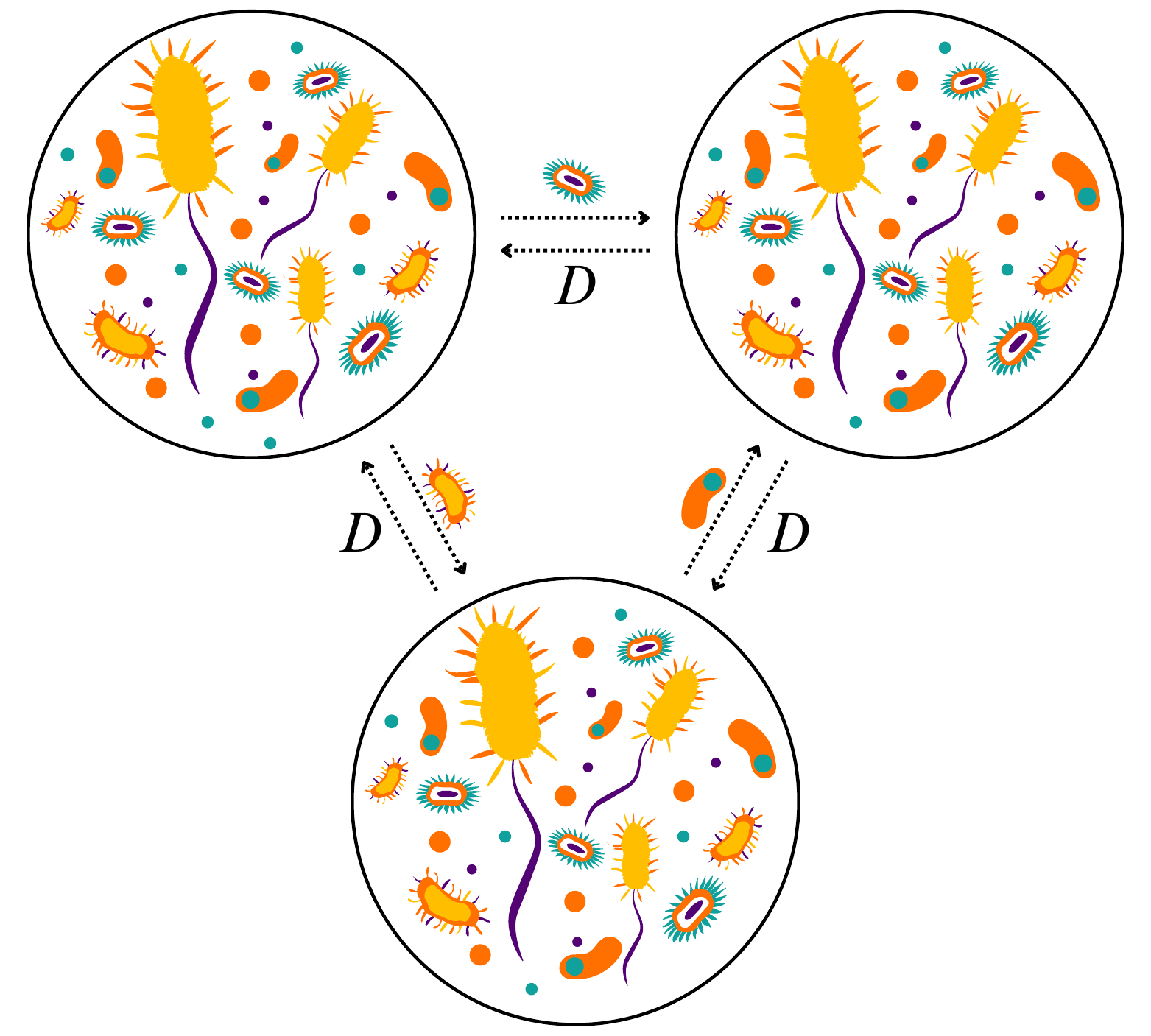}
\caption{A metacommunity of 7 species living on 3 patches. Each individual interacts with the local community to which it belongs possibly migrating to neighboring patches with diffusion coefficient $D$.}
	\label{fig:metacommunity}
\end{figure}
We consider a meta-community of $S$ species living on a network of $L$ discrete spatial locations, or patches.
A graphical representation of the system is given in Figure~\ref{fig:metacommunity} in the case of a fully connected network of 3 patches.
Each species is characterized by its abundance in each patch, which is modeled by a continuous variable, $N_{i,u}$, representing the total number of individuals divided by the typical size of the local population $\tilde{N}_{typ}$. 
The abundance of species $i$ in patch $u$ evolves according to the stochastic differential equation:
    \begin{align}
    \label{eq:LV}
    \begin{split}
    \dot{N}_{i,u}=\frac{r}{k}N_{i,u}\left(k-N_{i,u}-\sum_j\alpha_{ij}^u N_{j,u} \right)+\\\frac{D}{c} \sum_{v\in\partial u} \left(N_{i,v} -N_{i,u}\right) + \eta_i^u(t)\sqrt{N_{i,u}}  \,\,,      
    \end{split}
      \end{align}
which corresponds to Lotka-Volterra dynamics, with constant growth rate $r$ and carrying capacity $k$ that are set to 1 throughout. The notation $\partial u$ indicates the set of patch neighbors of $u$ (from and to which species in patch $u$ can migrate). 
The growth of each species is influenced by the abundance of all the others through the interaction coefficients $\alpha_{ij}^u$: if $\alpha_{ij}^u$ is positive species $j$ inhibits the growth of species $i$ in patch $u$ and vice versa. 
Positive $\alpha_{ij}^u$ and $\alpha_{ji}^u$ correspond to two species competing for resources, whereas  
$\alpha_{ij}^u$ and $\alpha_{ji}^u$ both negative correspond to mutualistic behaviour. Predation leads to opposite signs.

To model the heterogeneity in the interactions of species-rich ecosystems, we follow \cite{Kessler2015,bunin2016} and consider the disordered LV model. As already discussed in the introduction, the disorder approach has attracted recently a lot of attention \cite{bunin2017, biroli2018,altieri2021, altieri2022, larroya2023, galla2018, fisher2014} and also received positive experimental confirmations \cite{barbier2018,hu2022}. In this framework, 
the interaction coefficients are random variables, with mean $\mu/S$ and variance $\sigma^2/S$. They are independent in each patch except for $\alpha_{ij}^u$ and $\alpha_{ji}^u$, which have a correlation coefficient $\gamma$.
In the following, we will first focus on the symmetric interactions case ($\gamma=1$), and then show that a small asymmetry does not qualitatively change the results. 
As the interactions between species can depend on the environmental conditions (temperature, humidity, resources availability...) which differ in space, we consider interaction matrices fluctuating from one patch to another, i.e. they are not identical in different patches but corresponding elements $\alpha_{ij}^u$ and $\alpha_{ij}^v$ have a correlation coefficient $\rho$ \cite{Roy2020,pearce2020}.

We will restrict the choice of $\mu$ and $\sigma$ to values for which an isolated Lotka-Volterra community only displays a single uninvadable equilibrium (the single equilibrium phase studied in Ref. \cite{bunin2016}). 
Without spatial heterogeneity the transition point is not modified by the introduction of a spatial structure \cite{olmeda2023}, and spatial heterogeneity decreases the effective complexity of the interaction network \cite{gravel2016}, favoring the single equilibrium phase. Therefore we also expect the metacommunity to be in the single equilibrium phase for all the allowed values of $\mu$ and $\sigma$.
The effect of migration between patches in the strong heterogeneity regime with non symmetric interactions, in which a single  community with fixed immigration exhibits chaotic dynamics, \cite{bunin2017, biroli2018, altieri2021, hu2022, galla2018} was studied in \cite{Roy2020,pearce2020} in the absence of demographic noise. It leads to complex dynamical behavior with long-lived persistent fluctuations. Combining strong heterogeneity, demographic noise, and spatial migration is a challenge left for future studies. 

In the model defined by Eq.(\ref{eq:LV}) individuals can migrate on the patches network through diffusion, with a constant diffusion coefficient $D/c$, where $c$ is the connectivity (or number of connections per site) of the network. 
We assume the network to be translationally invariant, therefore each site has the same connectivity.
Migration is possible and equiprobable from patch $u$ to any of its $c$ nearest neighbors $v\in \partial u$. 

Each species is subject to a white demographic noise $\eta_i^u$, accounting for the stochasticity in birth and death events in a continuum setting \cite{altieri2021, larroya2023}.
We follow Ito's convention, according to which fluctuations in birth and deaths at time $t+dt$ depend on the abundance at the previous time step.
The noise is uncorrelated and of constant amplitude across species and patches:
\begin{align}
    \langle \eta_i^u(t) \eta_j^v(t')\rangle = 2 T \delta_{ij}\delta_{uv}\delta(t-t') \ . \end{align}
The auto-correlation of the demographic noise defines the noise strength $T$ which depends on the birth and death rates and on the typical size of the local population; 
$T$ scales as $T\propto 1/\tilde{N}_{typ}$ \cite{altieri2021, larroya2023}: the larger the local populations, the more negligible are demographic fluctuations.  In the $\gamma=1$ case $T$ can be interpreted as an effective temperature, as we shall show later. 

Some further insights into the effect of the demographic noise can be obtained considering it in the absence of all the other terms. 
In this case, an exact solution to the associated Fokker-Planck equation is available, showing that starting from any initial condition the population goes to 0 abundance with some finite rate \cite{feller1951, dornic2005}. Therefore also in the continuous model, extinction is possible over finite times, and not only asymptotically as it would be the case for example with environmental noise. 

\begin{figure}[htbp]
\centering
\includegraphics[width=0.35\textwidth]{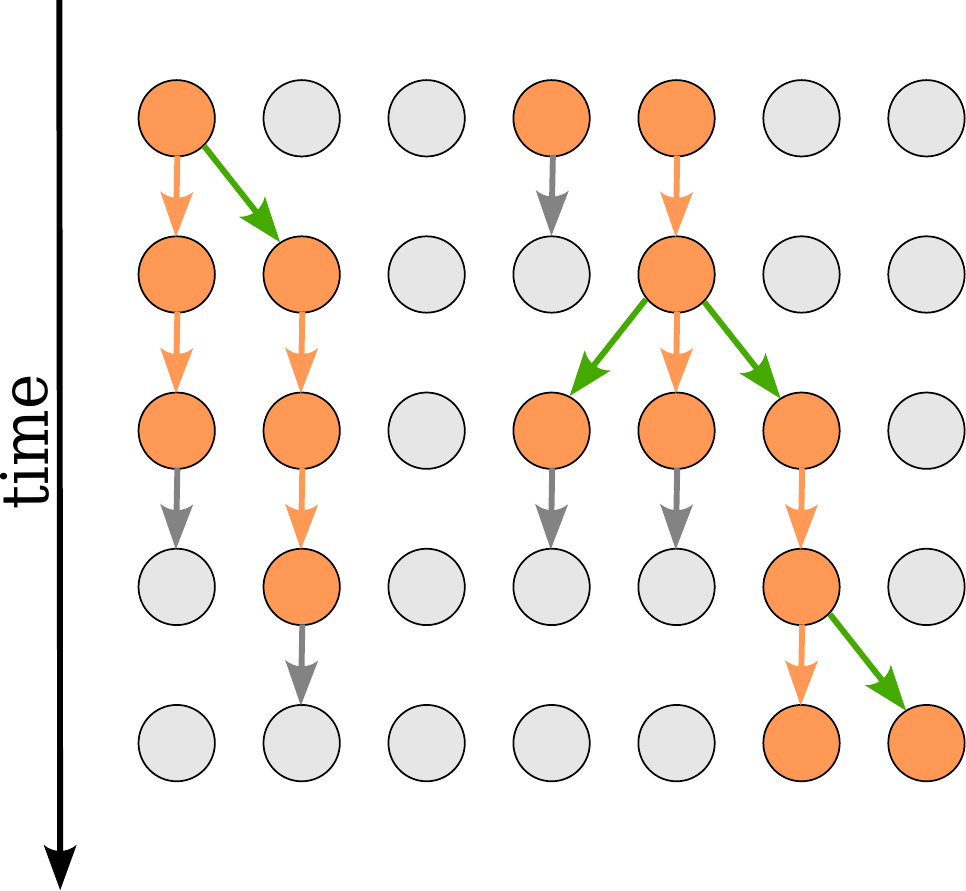}
\caption{Directed percolation on an array of 7 sites. Each row represents a different time step, green arrows indicate birth, gray arrows death, and orange arrows survival.}
	\label{fig:DP}
\end{figure}
\begin{figure}[htbp]
\centering
\includegraphics[width=0.35\textwidth]{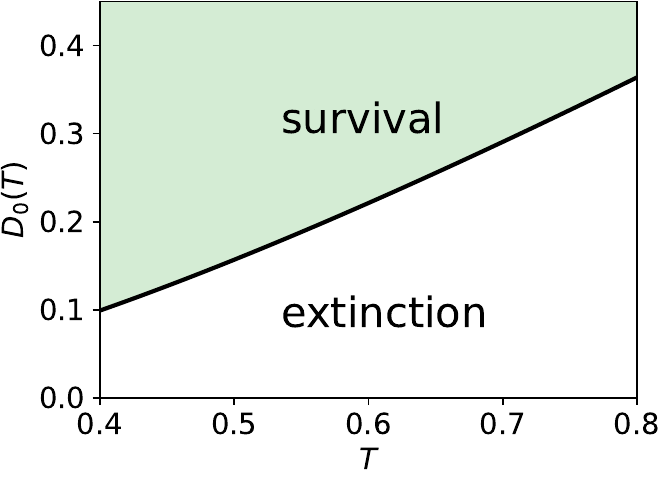}
\caption{Phase diagram for Directed Percolation in the mean-field approximation: in green the active phase, in which at long times there is a finite density of particles, in white the inactive phase, in which all particles eventually die. $D_0(T)$ indicates the transition line (see Sec. \ref{sec:methods} and App. \ref{app:dc} for details).}
	\label{fig:DPphasediagram}
\end{figure}

The dynamics of species in the presence of birth and death has important connections with the celebrated directed percolation problem studied in out-of-equilibrium physics and statistical field theory \cite{hinrichsen2000}.
Directed percolation is a model of particles that hop on a network and are subjected to births and deaths; a graphical illustration of the process can be found in Figure \ref{fig:DP} for a one-dimensional network. Directed percolation was originally introduced to model spreading phenomena, from forest fires to epidemics \cite{hinrichsen2000}. 
In our case, the sites of the network represent spatial locations, or patches, on which (or from which) species can migrate; the particles indicate which sites are colonized by species.
At each time step the particles can produce an offspring in a neighboring site, die or just survive. In our case, this corresponds to colonization or extinction.  
Depending on the competition between death and birth rates, the activity can spread to the entire system and lead to a finite density of particles (active, self-sustaining state) or die out (absorbing, inactive state). 
Between these two phases, there is a continuous phase transition, characterized by universal critical behavior \cite{cardy1980, hinrichsen2000}. We show in figure \ref{fig:DPphasediagram}) the phase diagram in the mean-field approximation (discussed in the next section).
A direct link between DP and GLV is obtained by coarse-graining \cite{cardy1980, hinrichsen2000}. In this way the discrete DP occupation variable becomes a continuous quantity that represents the mean occupation, the competition between birth and death rates gives rise to a logistic growth, hopping is replaced by diffusion and the stochastic fluctuations generate the demographic noise. This leads to a set of independent GLV eqs. (\ref{eq:LV}) in absence of interactions, one for each species. 
Each equation corresponds to an independent directed percolation process. 

The directed percolation transition can therefore be interpreted as a transition between a self-sustained phase where migration enables a finite abundance of species to persist to a regime, characteristic of small (or zero) dispersal, where species go extinct due to demographic noise. The aim of this work is to develop a theory for these phenomena for species-rich ecosystems in the presence of heterogeneous interactions. Upon increasing the number of species in the pool and considering heterogeneous interactions, the set of directed percolation processes is no longer independent and the complexity of the model increases considerably. In fact,  the system becomes equivalent to the collection of an infinite number of directed percolation processes, coupled by random interactions -- an interesting and open statistical physics problem.

\section{Methods} 
\label{sec:methods}

\subsection{DMFT and coupled Directed Percolation processes}

In this work, we aim to study systems in which both the number of species and the number of patches are very large. In order to obtain analytical results we follow the statistical physics "way" and take the limit 
of an infinite number of species and an infinite number of patches. 
In this double limit (whose order is irrelevant, see the appendix) the macroscopic properties of the system do not depend on the particular realization of the demographic noise and of the interactions: the macroscopic properties are \emph{self-averaging} in the jargon of disordered systems \cite{mezard1986}.

The large $S$ limit allows for an analytical treatment, as the dynamics of the $S$ interacting degrees of freedom can be replaced by the effective dynamics for a single representative species, through Dynamical Mean Field Theory (DMFT) \cite{roy2019}. 
The interaction effect with other species is captured by a noise term, which can be seen as an environmental noise (or a thermal bath) statistically defined in a self-consistent way.
The DMFT procedure is analogous to the one used to derive Langevin's equation from Newtonian dynamics \cite{zwanzig2001}, with the difference that here the degrees of freedom that are integrated out, giving rise to the noise, are equivalent to the degree of freedom under consideration, thus allowing a self-consistent closure of the equations of motion. DMFT is a very powerful technique that has been employed in several different contexts from quantum many-body systems to glassy dynamics \cite{georges1996,cugliandolo2023}. Thanks to DMFT, we can map an infinite number of randomly coupled DP processes -- a formidable problem -- into a {\it single} DP process with additional terms to be determined self-consistently (a colored noise and a memory term).

Our derivation follows the one developed in reference \cite{roy2019} for LV models, and can be found in Appendix \ref{app:DMFT} for generic values  $\rho$ of the spatial heterogeneity of the interactions.
Here we outline the main steps in the special case of patch-independent interactions, $\rho=1$. 
In the following, we are interested in the steady states of the dynamics. In fact, we expect that after a transient the system will settle in a time translationally invariant regime. 
For $S \to \infty$ DMFT allows one to replace the interaction term $-\sum_j\alpha_{ij}N_{j, u}$ by a stochastic expression that has the same statistical properties:
\begin{align}
   - \mu h-\sigma \tilde{\xi}^i_u(t)+ \sigma^2 \gamma \int_0^t \sum_v R_{uv}(t,s)N^i_v(s)ds \ . \end{align}
Since this allows us to decouple different species, we will for simplicity omit the species index $i$ in the following.
We now discuss the different contributions. Note that in the following empirical averages over species will be denoted as $\mathbb{E}[\cdot]$.

The first term represents the average interaction with all other species. It is given by the product of the mean of the interaction strength and the mean abundance, $h=\mathbb{E}[N_u]$, that in the steady state does not depend on the patch $u$ thanks to translational invariance. 

The second term represents the fluctuation of the interaction with all other species. It is given by the product of the standard deviation of the interaction coefficients and Gaussian noise with zero mean and correlation matching the time auto-correlation of the single species abundances:
\begin{align}
    \langle \tilde{\xi}_u(t)\tilde{\xi}_v(s)\rangle=\mathbb{E}[N_u(t)N_v(s)] :=C_{u,v}(t-s) \ .
\end{align}
The noise $\tilde{\xi}_u(t)$ is multiplied by the abundance in the LV equations. Henceforth we will call it \emph{environmental} since its effect is to add fluctuations to the carrying capacity. 
Since the autocorrelation of the abundances generically decays to a positive plateau at large time separations \cite{roy2019}, one can decompose the environmental noise into a fluctuating component and a static one. The former corresponds to the fluctuations due to ecological dynamics for a given species. The latter is characteristic of a given species and fluctuates from species to species \cite{roy2019}. 
We decompose the noise by rewriting $\tilde{\xi}_u(t)=z\sqrt{C_d^\infty}+\xi_u(t)$, where $C_{d}^\infty=\lim_{\tau\to\infty}C_{u,u}(t, t+\tau)$ is the value of the correlation function within the same patch at infinite times, $z$ is a static Gaussian variable with zero mean and unit variance, that now plays the role of quenched disorder, and $\xi_u(t)$ is a fluctuating noise whose covariance vanishes at long times. Again $z$ and $C_d^\infty$ do not depend on the patch $u$ thanks to translational invariance.

To distinguish the roles of fluctuating and static noises in the GLV equation,  we introduce two kinds of averages: $\langle \cdot \rangle $ indicates the average over the fluctuating noises $\xi$ and $\eta$. It is an average over the ecological dynamics, or by ergodicity, over patches for a fixed species. In analogy with physical systems, we call it \emph{thermal average}. The overline $\overline \cdot $ instead stands for the average over the static field $z$ corresponds to averaging over species or over different instances of the interaction matrix. Again in analogy with the physical system, we call it \emph{quenched disorder average}.

The last term in the dynamical mean-field treatment of the interactions is due to a feedback mechanism: a fluctuation of the abundance of species $i$ influences species $j$, which in turn influences species $i$.
These contributions sum up because of the correlation between $\alpha_{ij}$ and its reciprocal $\alpha_{ji}$, leading to the factor $\gamma$. 
This feedback mechanism (the famous Onsager reaction in the spin-glass literature) generates a memory term, containing the response function of the abundance on patch $u$ to a perturbation in the carrying capacity in patch $v$:
\begin{align}
    R_{u,v}(t,s)=\mathbb{E}\left[\frac{\delta N_u(t)}{\delta \zeta_v(s)}\bigg|_{\zeta=0}\right] \ .
\end{align}

In the $S\to \infty$ limit, there is convergence in law between the statistics of the infinite number of randomly coupled DP processes and the effective one \cite{arous1997symmetric,zwanzig2001}, i.e. the dynamics of a species satisfying the GLV equation (\ref{eq:LV}) is equivalent to the effective one of a {\it single species} living on the original spatial network:
\begin{widetext}
\begin{align}
\label{eq: DMFTspace}
    \dot{N_u}=N_u\left(k-N_u-\mu h-\sigma \left(z\sqrt{C_{d}^\infty}+ \xi_u\right)+ \sigma^2\gamma\int_0^t \sum_v R_{uv}(t,s)N_v(s)ds  \right)+ \frac{D}{c} \sum_{v\in\partial u} \left(N_v -N_u\right)+ \eta_u\sqrt{N_u} \ .
\end{align}
\end{widetext}
The DMFT closure consists then in replacing the empirical averages over species $\mathbb{E}[\cdot]$ with the one with respect to the effective single-species one. 
Because the effective process itself depends on some averaged quantities, one ends up with a self-consistent stochastic equation.


Eq. (\ref{eq: DMFTspace}) can also be interpreted as the Langevin equation associated with a Directed Percolation (DP) process, with the addition of a memory term (that is absent in the special case $\gamma=0$) and environmental noise. 
The effect of the static part of the environmental noise $z$ is to change the control parameter of the DP process, determining whether this is sub-critical or supercritical. 

Interestingly, whereas a system of few species interacting and diffusing on a network was established to boil down to a standard DP problem \cite{chen2016, janssen1997, mobilia2007, dobramysl2018, broadbent1957}, the case of {\it many} species is fundamentally different and belongs to a different class. 
Indeed, a system of {\it many} species is equivalent to a family of {\it many} DP processes, characterized by different values of static and fluctuating noises and coupled through the common self-consistently determined mean, correlation, and response functions.
Understanding the behavior of this self-consistent DP problem is an open challenge. In this work, we study whether the DP transition can fundamentally change nature due to this self-consistent coupling. Even if the transition remained qualitatively DP-like (continuous and from an absorbing state to a fluctuating one) critical properties could change. In fact, although an environmental noise can be shown to be an irrelevant perturbation of the associated field theory \cite{hinrichsen2000}, within DMFT the environmental noise inherits the time dependence of the correlation function through the self-consistency. 
It can therefore develop long-range correlations in time at the critical point, possibly altering the critical behavior and leading to a new universality class.

\subsection{Symmetric interactions, mean-field approximation, and mapping to a system in thermal equilibrium}

Studying the coupled field theories (\ref{eq: DMFTspace}) is a formidable task. In the following, we simplify the problem by doing a mean-field approximation which allows us to obtain a general theory independent of the underlying network of patches. 

We replace the term 
$\frac{D}{c} \sum_{v\in\partial u} N_v$ by its thermal average. This amount to $\frac{D}{c} \sum_{v\in\partial u} N_v \rightarrow DN^*$, where $N^*=\frac{1}{c}\sum_{v\in\partial u}\langle N_v\rangle$ and, using translation invariance, it simplifies to $\langle N_u\rangle$ (which is time-independent since we are considering steady states). This procedure corresponds to a mean-field approximation of the spatially dependent DMFT Eqs. (\ref{eq: DMFTspace}). Such DMFT$^2$ approximation becomes exact for a fully connected network.  In fact, in this case, taking the $L\to\infty$ limit, the empirical average of the abundances over the patches concentrates around the thermal average  $N^*=\langle N_u\rangle$. 
From now on, we shall focus on this case. 

By substituting $\frac{D}{c} \sum_{v\in\partial u} N_v$ with $N^*$ in Equation (\ref{eq: DMFTspace}), one obtains an equation on $N_u$ only, 
with an additional parameter to be determined self-consistently. 
Note that $N^*$ is obtained by averaging only over thermal fluctuations, and not over disorder: therefore, it will have to be determined as a function of $z$. 
This means that different species will have different immigration rates (here, for simplicity, we are still focusing on the $\rho=1$ case; generalizations will be discussed later).

This substitution allows us to decouple stochastic processes for the abundance in different patches. Omitting for simplicity the index $u$, we now obtain (for large times, i.e. in the steady state):
\begin{widetext}
\begin{align}
\begin{split}
    \dot{N}=N\Bigg(k-N-\mu h- \sigma z\sqrt{C_d^\infty}-\sigma \xi(t)+\sigma^2\gamma\left( \int_0^t  R_d(t-s)N(s)ds+N^*\int_0^t  R_0(t-s)ds\right)\Bigg)+\\+ D (N^*-N) + \eta(t)\sqrt{N}
\end{split}
\end{align}    
\end{widetext}

Since all patches are equivalent on a fully connected lattice, the $R_{uv}$ and $C_{uv}$ matrices (of functions) only have two independent elements: the diagonal ones, $R_d$ and $C_d$, and the off-diagonal ones $R_0/L$ and $C_0$ 
(see the appendix for the justification of the scaling with $L$ of $R_0/L$ and $C_0$).

In the case of symmetric interactions, $\gamma=1$,
one can show (see App. \ref{app:Hamiltonian}) that the self-consistent solution maps to a thermal equilibrium process. In fact, one finds that the diagonal elements of the response and correlation functions obey a fluctuation-dissipation relation:
\begin{align}
    R_d(\tau)=-\frac{1}{T}\frac{\partial }{\partial \tau}C_d(\tau) \ .
\end{align}
The memory term and $\xi$ therefore play the role of a friction term and the noise associated with a colored thermal bath at temperature $T$. 
The stochastic process maps then to a generalized Langevin equation whose stationary probability distribution is given by the Boltzmann distribution at temperature $T$ and with the effective Hamiltonian:
\begin{align}
\begin{split}
	H_{eff} = \left( 1-\frac{\sigma^2}{T}(C_d^0-C_d^\infty) \right)\frac{N^2}{2} -\big(k-D -\mu h+\\- z \sqrt{C_d^\infty}\sigma + \sigma^2 N^* R_0^{int}  \big)N + (T-D N^*)\ln N \ ,
\end{split}
\end{align}
where $C_d^0$ is the equal-time correlation function, namely the second moment of the abundances over disorder and noise, $\overline{\langle N^2\rangle}$. 
The long time limit of the correlation function, $C_d^\infty$, represents instead the second moment of the thermal-averaged abundances, $\overline{\langle N\rangle^2}$.
$R_0^{int}$ is the integrated off-diagonal response, which is the solution of the self-consistent equation (see Appendix \ref{app:responsefunction}):
\begin{align}
    R_0^{int} =\overline{r_d(z)\frac{D \chi(z) +\sigma^2 R_0^{int} r_d(z)}{1-\left( D \chi(z)+\sigma^2R_0^{int}r_d(z)\right)} }\ .
\end{align}
$\chi(z)$ and $r_d(z)$ are the species-dependent response to a perturbation in the immigration rate or the carrying capacity, respectively:
\begin{align}
    \chi(z)&=\langle N \log N \rangle -\langle N \rangle \langle \log N\rangle\\
    r_d(z)&=\langle N^2\rangle-\langle N\rangle ^2
\end{align}

The self-consistent equations can be expressed as averages with respect to the Boltzmann distribution:
\begin{align}
    N^*(z)&=\langle N \rangle= \frac{\int_0^\infty dN N e^{-\beta H_{eff}}}{\int_0^\infty dN e^{-\beta H_{eff}}}\label{eq:selfconsN*}\\
    h&=\overline{\langle N\rangle}=\int \mathcal{D} z \frac{\int_0^\infty dN N e^{-\beta H_{eff}}}{\int_0^\infty dN e^{-\beta H_{eff}}}  \\
    C_d^0&=\overline{\langle N^2\rangle}=\int \mathcal{D} z \frac{\int_0^\infty dN N^2 e^{-\beta H_{eff}}}{\int_0^\infty dN e^{-\beta H_{eff}}}\\
    C_d^\infty&=\overline{\langle N\rangle^2}=\int \mathcal{D} z \left(\frac{\int_0^\infty dN N e^{-\beta H_{eff}}}{\int_0^\infty dN e^{-\beta H_{eff}}}\right)^2\label{eq:selfconsq0} \ .
\end{align}
and analogously for $R_0^{int}$. 
$\int \mathcal{D} z=\int \frac{d z}{\sqrt{2 \pi}} e^{-z^2/2}$ indicates the average over the Gaussian field.

These equations can be solved iteratively: starting from a suitable initial condition for $N^*(z)$, $h$, $C_d^0$, $C_d^\infty$ and $R_0^{int}$, one updates their values according to equations (\ref{eq:selfconsN*})-(\ref{eq:selfconsq0}) until reaching a fixed point. 
Because very large values of $z$ are exponentially suppressed by the Gaussian distribution, it is sufficient to determine $N^*(z)$ for $z$ of $O(1)$. 

In conclusion, within the DMFT$^2$ approximation and for the symmetric case, the formidable self-consistent stochastic equations (\ref{eq: DMFTspace}) can be analyzed by studying a set of static self-consistent equations on four parameters $h,C^0_d,C^\infty_d, R_0^{int}$ and one function $N^*(z)$.
Solving these equations (see next section) allows us to obtain a general picture of the interplay between migration and demographic noise for spatially extended metacommunities. In order to show that such a picture is valid beyond the simplified case we focus on, we have also considered several extensions that we shall present below.


\subsection{Extensions}
\subsubsection{Spatial heterogeneity}

In the case of a generic value of the spatial heterogeneity of the interactions $\rho$, an analogous procedure can be implemented, with some important differences.
The static disorder is now a patch-dependent and correlated variable, that we can decompose as
$\rho \sqrt{C_0^\infty} z+ \sqrt{C_d^\infty-\rho^2 C_0^\infty} w_u$ where $z$ is constant and $w_u$ independent across locations, and $C_d^\infty$ and $C_0^\infty$ are the infinite time correlation function of the abundance on the same patch and on different patches, that for $\rho=1$ coincide.
Averaging the abundance across patches to obtain the immigration rate requires an additional step, \emph{i.e.} averaging also over $w_u$.
The solution of the self-consistent equations, albeit conceptually analogous to the $\rho=1$ case, is for generic values of $\rho$ much more numerically challenging, because of the need to integrate over two disorder fields, $z$ and $w_u$. 
For this reason, we focused on the two extreme cases, $\rho=1$ and $\rho=0$, in which only one of the two disorder fields is present. The results are qualitatively similar so we expect our conclusions to hold also for intermediate values of $\rho$. We confirm it by numerical simulations at $0<\rho<1$.

\subsubsection{Asymmetric interactions}
The mapping to an equilibrium distribution requires symmetry in the interactions: non-symmetric interactions correspond to non-conservative forces, which explicitly break time reversal and lead to non-equilibrium steady states.
In order to show that our results hold also in this case, at least if the asymmetry is not too strong, we have analyzed the case of small asymmetry in perturbation theory. 
The analysis of the Martin-Siggia-Rose-De Dominicis-Janssen action \cite{martin1973, janssen1976, dedominicis1978, aron2010} allows us to conclude that a small degree of asymmetry ($\gamma=1-\epsilon$, $\epsilon \ll 1$) does not affect qualitatively the results we shall present in the next section, therefore establishing that our findings for the symmetric case also holds for small asymmetry (see Appendix \ref{app:asymmetry} for more details). We have also confirmed this result by numerical simulations for $\gamma<1$.

\section{Results} 
\label{sec:anresults}
In the following we present our analytical results focusing on ecosystems with parameters $\sigma=0.5$ and $\mu=1$, hence a case in which interactions are in average competitive for the pool of species. 
\begin{figure}[htbp]
\centering
     \begin{subfigure}[b]{0.23\textwidth}
         \centering
         \includegraphics[width=\textwidth]{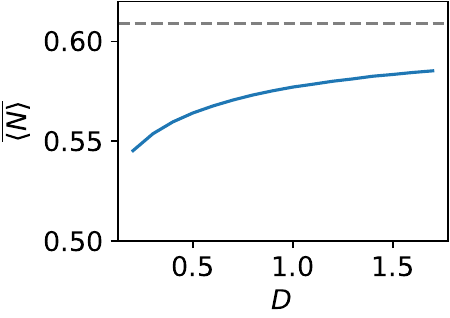}
     \end{subfigure}
     \hfill
\begin{subfigure}[b]{0.23\textwidth}
         \centering
         \includegraphics[width=\textwidth]{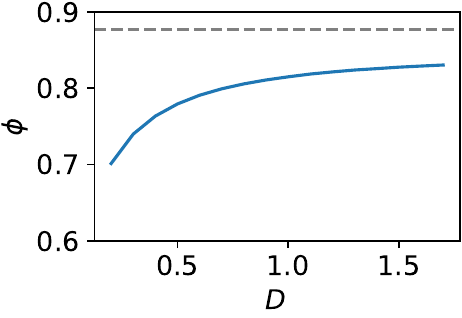}
     \end{subfigure}
     \hfill
     \begin{subfigure}[b]{0.23\textwidth}
         \centering
         \includegraphics[width=\textwidth]{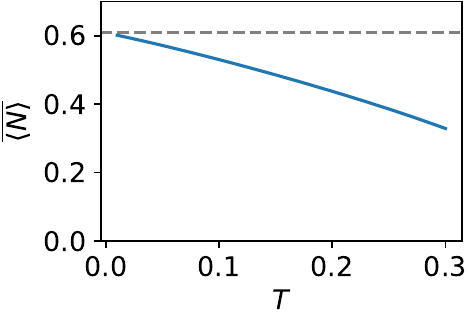}
     \end{subfigure}
     \hfill
\begin{subfigure}[b]{0.23\textwidth}
         \centering
         \includegraphics[width=\textwidth]{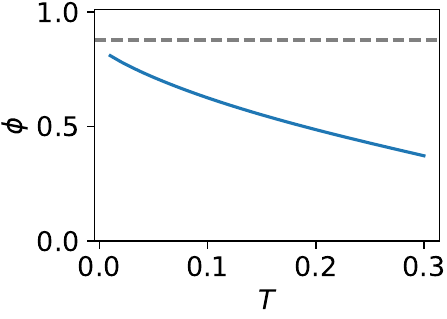}
     \end{subfigure}
     \hfill
\caption{ Average abundance $\overline{\langle N\rangle }$ and diversity $\phi$ as a function of the diffusion constant $D$ for $T=0.25$ (top) and as a function of temperature (strength of demographic noise) for $D=0.1$ (bottom). 
The dashed lines represent the $T=0$ well-mixed results. $\mu=1$, $\sigma=0.5$.
}
	\label{fig:Dlarge}
\end{figure}

\begin{figure*}
\centering

\sbox{\bigpicturebox}{%
  \begin{subfigure}[b]{.47\textwidth}
  \scalebox{1}[1]{\includegraphics[width=\textwidth]{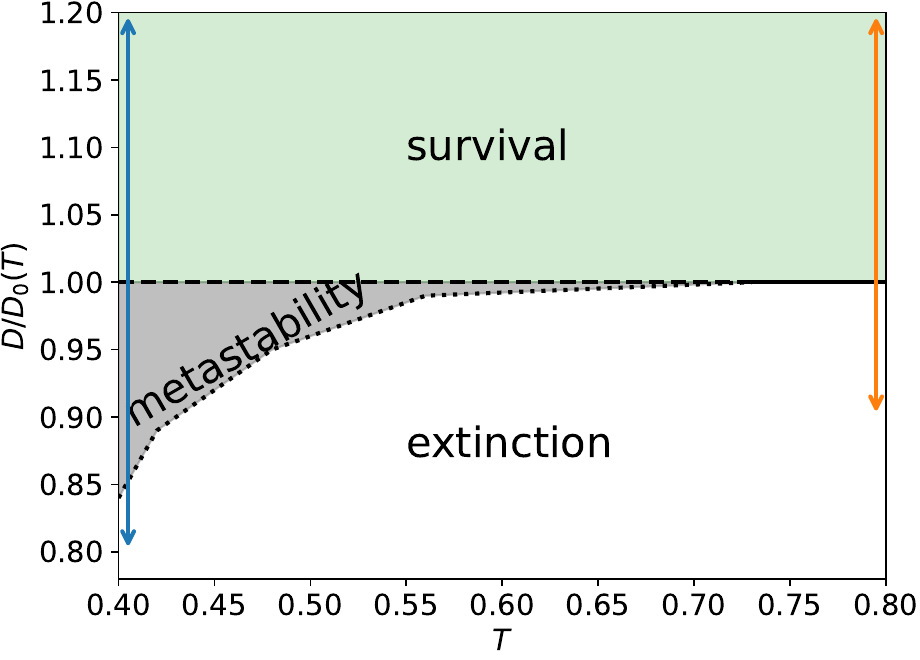}}%
\caption{Phase diagram, $\rho=1$}
\label{fig:phasediagramrho1}
\end{subfigure}
}

\usebox{\bigpicturebox}\hfill
\begin{minipage}[b][\ht\bigpicturebox][s]{.47\textwidth}
\begin{subfigure}{.49\textwidth}
\includegraphics[width=\textwidth]{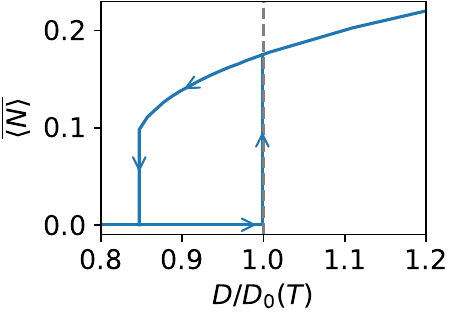}
\caption{$T=0.4$}
\label{fig:hvsDdiscont}
\end{subfigure}\hfill
\begin{subfigure}{.49\textwidth}
\includegraphics[width=\textwidth]{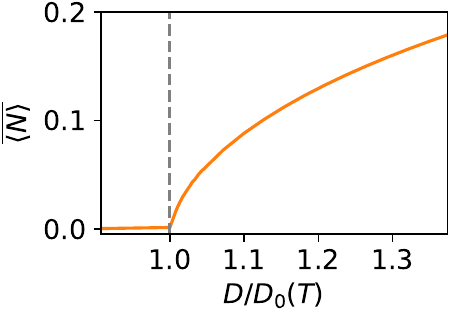}
\caption{$T=0.8$}
\label{fig:hvsDcont}
\end{subfigure}

\begin{subfigure}{.49\textwidth}
\includegraphics[width=\textwidth]{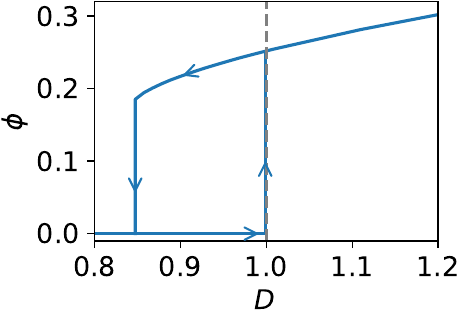}
\caption{$T=0.4$}
\label{fig:phivsDdiscont}
\end{subfigure}\hfill
\begin{subfigure}{.49\textwidth}
\includegraphics[width=\textwidth]{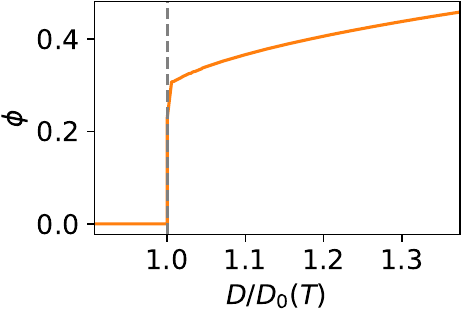}
\caption{$T=0.8$}
\label{fig:phivsDcont}
\end{subfigure}

 \end{minipage}
\caption{(a) The phase diagram for constant interactions across patches ($\rho=1$). The continuous line indicates the continuous transition, and the dotted and dashed lines are the limits of the metastability region, highlighted in grey. At the two limits of the metastability region one of the two solutions disappears and we have a discontinuous transition. The arrows indicate the parameters range in the right figures.
The average abundance $h=\overline{\langle N\rangle}$ and the diversity $\phi=\overline{\theta(\langle N\rangle)}$ as a function of $D$ across a discontinuous (b, d) or continuous (c, e) transition. In subfigure (c) the arrows indicate the direction of the hysteresis cycle: decreasing $D$ (starting from high values) the ecosystem would follow the finite solution until the discontinuous transition, where the abundances jump to zero. If we now increase $D$, it would follow the zero solution until this becomes unstable at $D_0(T)$. Gray dashed lines indicate the value of $D$ at which a single species would go (continuously) extinct. Note that we have divided $D$ by the critical value of the diffusion constant for Directed Percolation $D_0(T)$ in all plots, to emphasize the effect of interactions on the already known case. Because $D_0(T)$ vanishes exponentially for $T\to 0$ (App. \mbox{\ref{app:dc}}), the metastability region has a vanishing width in this limit and the system is always in the survival phase.
$\mu=1$, $\sigma=0.5$.}
\label{fig:rho1}
\end{figure*}
\subsection{Characterization of the self-sustained phase}

By solving the DMFT equations described in the previous section, one finds that when the diffusion constant is large enough the system is in a self-sustained phase (active phase in the directed percolation jargon) in which a non-zero abundance is maintained despite the presence of demographic fluctuations. In this regime, although some species go globally extinct on all patches, others survive thanks to the migration from neighboring patches. 
This mechanism is sufficient to prevent extinctions due to demographic stochasticity and leads to a self-sustained metacommunity. 

In the following, we discuss the salient properties of this phase, focusing on two ecologically relevant observables: the average abundance, $h=\overline{\langle N \rangle}$, and the ecosystem diversity $\phi$, defined as the fraction of species that are not globally extinct, \textit{i.e.} that have non zero abundance in at least one patch. 
At stationarity, we can compute the ecosystem diversity as $\phi=\overline{\theta(\langle N\rangle)}$. 

As expected, demographic noise is detrimental to survival: the fraction of surviving species, or diversity, and the average abundance decrease with the strength of demographic fluctuations, see bottom panels of Fig. \ref{fig:Dlarge}. On the contrary, dispersal is beneficial, as shown in the top panels of Fig. \ref{fig:Dlarge}.
The behavior of the diversity for species-rich ecosystems with heterogeneous interactions in the presence of demographic noise is a novel result of our approach: in the case of fixed external immigration, previously often considered in the literature, all species are kept alive by the immigration, albeit some at very small abundances, it is therefore not possible to rigorously define the ecosystem diversity \cite{garcialorenzana2022}. 
We find that the species that go extinct are those whose growth is on average more affected by the interactions with the rest of the ecosystem, as quantified by the static part of the environmental noise $z\sigma \sqrt{q_0}$, which renormalizes the carrying capacity of a species. 
For $\rho=1$, if $z$ is larger than a critical value $z^*$ 
the corresponding species goes extinct (for $z>z^*$ the renormalized carrying capacity is negative). This is true also for smaller values of $\rho$ (Appendix \ref{app:zstar}).
The case of independent interactions across patches ($\rho=0$) is special, for all species are globally equivalent so that they can only be all surviving or all extinct.
In general, all species have some patches in which they are very abundant, immigrants from these patches can then save them from extinction in the rest of the system.
This favorable role of dispersal through which spatial heterogeneity enhances diversity has been discussed in \cite{leibold2004, Roy2020, pettersson2021, gravel2016}.


The limits $D\to\infty$ and $T\to 0$ can be mapped to the well-mixed case. For $D\to\infty$ the timescale of spatial mixing is much smaller than all other timescales, therefore the abundances of each species are equal on all sites. 
The absence of spatial fluctuations allows one to write an evolution equation involving only the space-averaged abundances, that corresponds to an effective single local community without demographic fluctuations with interactions given by the spatial average of the original ones. 
The well-mixed result is also recovered (for $\rho=1$) in the $T\to0$ limit (see two bottom panels of Figure \ref{fig:Dlarge}): because the abundances do not fluctuate there is no migration flux between patches, and the diffusion term plays no role.

As for the distribution of the abundances, we find an exponential decay (see Appendix \ref{app:PN}), as it is the case in other models with random fully connected interactions \cite{bunin2017, biroli2018, altieri2021, wu_understanding_2021}.

\subsection{Transition to complete extinction: emergence of a discontinuous transition at low dispersal}

When demographic fluctuations are sufficiently strong, 
decreasing the diffusion constant leads to a continuous phase transition from an active phase in which some species are able to self-sustain to an inactive phase in which they are all extinct. 
The critical value of the diffusion constant is the same that would be obtained in the absence of interactions, where the system directly maps to directed percolation, or in the case of constant interactions \cite{denk2022}, see Figure \ref{fig:DP} and Appendix \ref{app:dc}. 
This is to be expected: upon approaching the transition, the abundances tend to zero, and therefore the interactions, which have a quadratic dependence on the abundances, become irrelevant.
The critical exponents indeed match the ones falling in the Directed Percolation universality class; in particular, the abundance goes to zero linearly (Figure \ref{fig:hvsDcont}). 
Interestingly, approaching the transition the diversity does not go to zero and instead tends to a finite value (Figure \ref{fig:phivsDcont}). The average abundance goes to zero not because more and more species are going extinct, but because all surviving species are simultaneously decreasing their abundances.
This homogenization in the behavior of species is yet another consequence of the irrelevance of the interactions, the only trait distinguishing one species from another in our model.

At smaller demographic noise this picture changes drastically and interactions play a major role, as shown in the phase diagram in Figure \ref{fig:phasediagramrho1}.
The ecosystem is able to self-sustain at values of the diffusion constant for which in the absence of interactions it would be in the inactive phase.
Further lowering $D$ we encounter a discontinuous transition at which all species abruptly go extinct, i.e. species abundances suddenly jump to zero. 
Before the discontinuous transition, there is an extended region in which the ecosystem is meta-stable (in grey in Figure \ref{fig:phasediagramrho1}): in this regime, the system reaches an equilibrium with high or low abundances depending on the initial conditions. It exhibits hysteresis (Figure \ref{fig:hvsDdiscont}). 

\begin{figure}[htbp]
         \centering
     \begin{subfigure}[b]{0.23\textwidth}
         \centering
         \includegraphics[width=\textwidth]{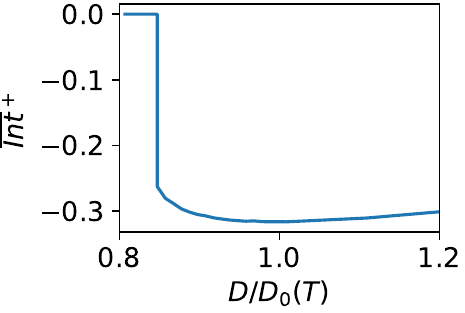}
         \caption{}
         \label{fig:intAn}
     \end{subfigure}
     \begin{subfigure}[b]{0.23\textwidth}
         \centering
         \includegraphics[width=\textwidth]{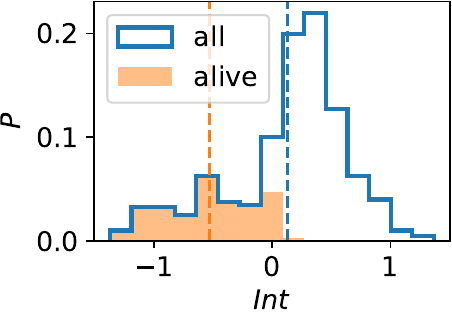}
         \caption{}
         \label{fig:intNum}
     \end{subfigure}
\caption{Thermal averaged interaction term, $Int=\langle\sum_j\alpha_{ij}N_j\rangle$, averaged over non extinct species (indicated by an overline with a $^+$ superscript), for two temperatures corresponding to the discontinuous regime. Left: analytical results for $T=0.4$, $\rho=1$ (as in Figures \ref{fig:rho1}b-d). $Int^+$ is negative in the metastability region, it jumps to zero when all species go extinct at the discontinuous transition. Right: Distribution of the thermal averaged interaction terms in a numerical simulation in the metastability region ($T=0.18$, $D/D_0(T)=0.8$, $S=200$, $L=400$, $t_{max}=500$, averaged over 2 runs). Non extinct species are highlighted in orange, only species with negative (or close to zero) interaction terms manage to survive. Averaging only over non extinct species (orange dotted line) leads to a significantly lower (more mutualistic) value than averaging over all species (blu dotted line). $\mu=1$, $\sigma=0.5$}
	\label{fig:interactions}
\end{figure}

It was recently shown that a metacommunity subject to demographic noise and constant mutualistic interactions exhibits a similar discontinuous phase transition \cite{denk2023}. 
The authors of \cite{denk2023} also performed numerical simulations with random (patch-independent) interactions, showing that the surviving species have more mutualistic interactions than the total species pool. 
We find that a similar mechanism is at play in our case: it is an emergent phenomenon due to ecological dynamics which is present even though interactions are not on average mutualistic (in fact they are competitive, $\mu=1$). Because of the symmetry in the interaction network, species that interact more competitively are more negatively affected by the interactions with the rest of the ecosystem, and will hence be more easily driven to extinction.
This leads to a decrease of the mean of the interaction matrix restricted to surviving species, which we have estimated in the case $\rho=1$ using a result obtained in \cite{baron2023} (Appendix \ref{app:restrictedmatrix}). 
Another quantity of interest is the average interaction term for non-extinct species, $\overline{Int}^+=\overline{\sum_j\alpha_{ij}\langle N_j\rangle}^+$ (the $^+$ indicates that the average is carried out only over non-extinct species, $\langle N_i\rangle>0$), which we find to be negative in the entire metastability region (Fig. \mbox{\ref{fig:intAn}}).
In order for a species to survive in conditions in which without interactions it would go extinct, we need the interaction term (that appears summed to the carrying capacity with a negative sign) to give on average a negative contribution. 
We indeed find numerically that only species with negative interaction terms manage to survive (Fig. \mbox{\ref{fig:intNum}}), thus leading to an enhancement of mutualism between surviving species -- see Appendix \mbox{\ref{app:restrictedmatrix}} for details.

\begin{figure}[htbp]
         \centering
         \includegraphics[width=0.47\textwidth]{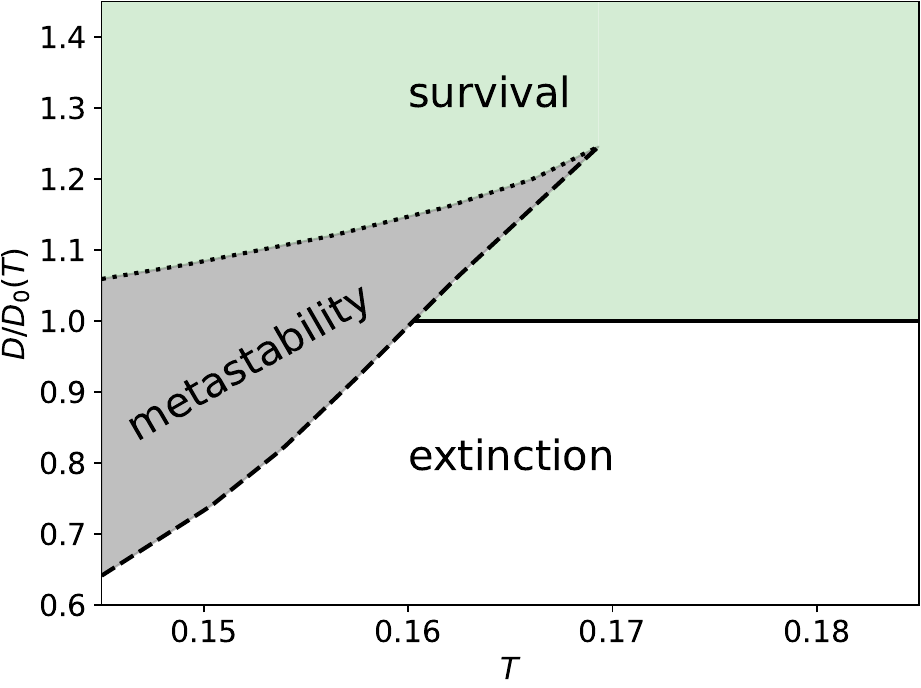}
\caption{The phase diagram for independent interactions across patches ($\rho=0$). The continuous line indicates the continuous transition, the dotted and dashed lines the limits of the metastability region, highlighted in grey. At the two limits of the metastability region one of the two solutions disappears and we have a discontinuous transition. $\mu=1$, $\sigma=0.5$}
	\label{fig:phasediagramrho0}
\end{figure}

In Figure \ref{fig:phasediagramrho0} we also show the phase diagram in the case of independent ($\rho=0$) interactions across patches, to be compared to the one of Figure \ref{fig:phasediagramrho1} corresponding to constant ($\rho=1$) interactions across patches. In both cases, the upper limit of the metastability region is bounded from below by the critical value of the diffusion constant in the absence of interactions, $D_0(T)$. 
For $\rho=1$ these two lines coincide, whereas for $\rho=0$ the metastability region extends above $D_0(T)$ in some range of temperature. 
In the part of the metastability region above $D_0(T)$ the two metastable solutions are both finite: one is of order one and the other proportional to the distance from $D_0(T)$; the two solutions coalesce at the tip of the metastability region.

One can also analytically show that the phase diagrams remains qualitatively unchanged considering a small asymmetry in the interactions ($\gamma=1-\epsilon$, $\epsilon\ll 1$), see Appendix \ref{app:asymmetry}. Numerical simulations presented in the next sections confirm this result.

\section{Assessing the generality of the scenario} 
To confirm the generality of our results, we now consider different variations of the model studied in the previous section. The aim is to show that our results hold in a broader setting. We shall be particularly interested in considering the case of a large but finite number of species, a large but finite number of patches, a small but finite asymmetry of interactions, as well as intermediate values of $\rho$. All these cases could be in principle studied analytically but they would require very involved (in some cases very challenging) analysis. 
We therefore turn to direct numerical simulations of the Generalized Lotka-Volterra equation \mbox{(\ref{eq:LV})} and show that the results agree with and extend the theory presented in the previous section. The details on the numerical scheme implemented for the simulation can be found in Appendix J.
These simulations are challenging as we are interested in considering both a large number of species and a large number of patches. Moreover, lowering the temperature results in a strong slowdown of the dynamics (Appendix \ref{app:numres}), leading to additional computational costs. The slowdown of the dynamics is much stronger in the presence of heterogeneity in the interactions than with zero or constant ones.


\subsection{Finite number of species and finite number of patches}
\label{sec:numresults}

\begin{figure}[htbp]
\centering
     \begin{subfigure}[b]{0.23\textwidth}
         \centering
         \includegraphics[width=\textwidth]{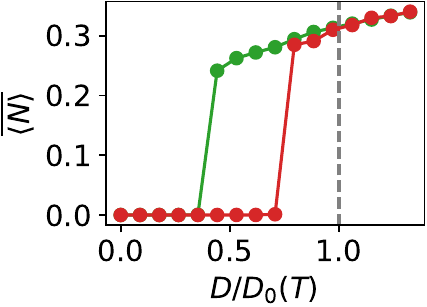}
         \caption{$T=0.18$}
         \label{fig:hT02}
     \end{subfigure}
     \hfill
\begin{subfigure}[b]{0.23\textwidth}
         \centering
         \includegraphics[width=\textwidth]{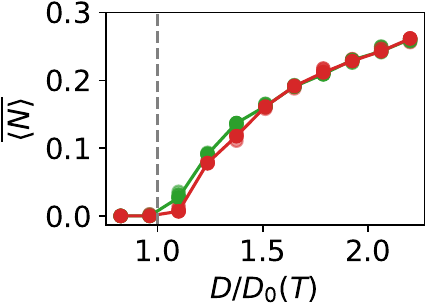}
         \caption{$T=0.8$}
         \label{fig:hT04}
     \end{subfigure}
     \hfill
\caption{Average abundance $\overline{\langle N\rangle }$ as a function of the diffusion constant $D$ for $T=0.18$ and $T=0.8$. Green and red dots indicate the initial conditions of order $1$ and of order $0.1$. The dashed line indicates the analytical prediction for the critical value of the diffusion constant for the continuous transition. $\mu=1$, $\sigma=0.5$, $S=200$, $L=400$, $t_{max}=500$ (left) and $200$ (right).
}
	\label{fig:numericalh}
\end{figure}

Generically, for moderate system sizes ($S<100$ and $L<100$) we find strong fluctuations due to the quenched disorder in the interaction matrix, and quantitative finite size effects compared to the asymptotic $S,L\rightarrow \infty$ solution, in particular for $\rho=1$ (for $\rho=0$ each patch is characterized by an independent realization of the interaction matrix, thus leading to a faster (self-averaging) convergence of the system to its disorder average). For larger values of $S$ and $L$, e.g. $S=200$, $L=400$, fluctuations and finite size effects are limited and one finds results that are both qualitative and quantitative in agreement with the analytical solution. 

In figure \ref{fig:numericalh} we show the behavior of the average abundance as a function of the diffusion constant for two different values of the temperature, starting from two different initial conditions. In order to probe the existence of hysteresis, and therefore a discontinuous transition and metastability, we numerically simulate systems with different initial conditions. For the green curves, the initial abundances were uniformly sampled between 0 and 1, for the red curves between 0 and 0.1. The former should therefore be more prone to evolve toward the self-sustained solution, if it exists, whereas the latter to the "all-extinct" solution. 

We find that indeed at higher temperatures, $T=0.8$, in agreement with the analytics and the phase diagram in Figure \ref{fig:phasediagramrho1}, the final abundances vary continuously when varying the diffusion constant, and they converge to the same value, no matter the initial condition.
The value of $D$ at which the final abundances significantly depart from zero quantitatively matches the analytical result for the critical value of the diffusion constant at the continuous transition. 

Instead, at $T=0.18$ the final abundances show a strong dependence on the initial condition in an extended interval of diffusion strengths; for a given initial condition the final abundance exhibits a very abrupt change \footnote{
The discontinuous transition takes place slightly before the analytical prediction. Besides finite size effects, we note that this phenomenon is to be expected for this kind of transition. In fact, 
when the red curve (low initial condition) jumps to high abundance, this does not necessarily indicate that the $\overline{\langle N\rangle}=0$ solution has become locally unstable, but rather that its basin of attraction has shrunk and does not include the considered initial condition anymore. 
It is therefore to be expected that this occurs for $D<D_0(T)$. A similar phenomenon takes place for spinodal transition in physics.}.
Interestingly, the dynamics strongly slows down in this regime, in particular for the decay of the abundances from large initial conditions. In fact, this process occurs via the rare extinctions of species that are asymptotically not able to self-sustain but can persist for very long times, especially in this regime in which demographic fluctuations are weak. 
The strong dependence on the initial conditions cannot be explained just by the slowdown of the dynamics because the abundances with different initial conditions evolve in opposite directions (see Figure \ref{fig:numericalh} and Appendix \ref{app:numres}). 

The heterogeneity in the interaction network is essential to allow the ecosystem to self-sustain below the single DP critical point: indeed if we consider the same parameters but take $\sigma=0$ all species go extinct below $D_0(T)$, and there is no strong dependence on the initial conditions (Appendix \ref{app:numres}). 

\subsection{Asymmetric interactions and partial correlation between patches}
\begin{figure}[htbp]
\centering
     \begin{subfigure}[b]{0.23\textwidth}
         \centering
         \includegraphics[width=\textwidth]{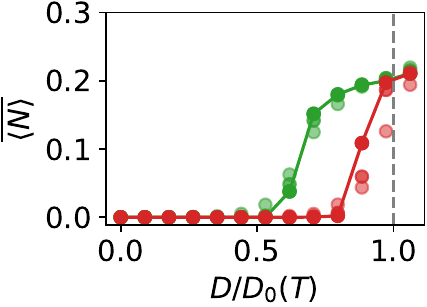}
         \caption{$T=0.18$}
         \label{fig:hT02asym}
     \end{subfigure}
     \hfill
\begin{subfigure}[b]{0.23\textwidth}
         \centering
         \includegraphics[width=\textwidth]{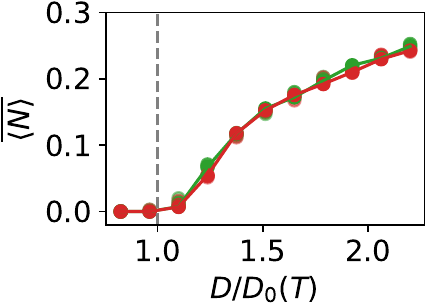}
         \caption{$T=0.8$}
         \label{fig:hT04asym}
     \end{subfigure}
     \hfill
\caption{Average abundance $\overline{\langle N\rangle }$ as a function of the diffusion constant $D$ for $T=0.18$ and $T=0.8$ with some spatial heterogeneity ($\rho=0.9$) and some asymmetry in the interactions ($\gamma=0.9$). Green and red lines indicate the initial conditions of order 1 and of order 0.1, lighter dots show the average abundance at intermediate times ($50\%$ and $75\%$ of $t_{max}$).  $\mu=1$, $\sigma=0.5$, $S=200$, $L=400$, $t_{max}=500$ (left) and $200$ (right).}
	\label{fig:numericalhasym}
\end{figure}
We are now interested in focusing on cases in which the interactions between species are not fully symmetric, and the interaction matrices are partially correlated between patches, i.e. $0<\rho<1$. 

As we have already discussed, we have analytically established that a very small asymmetry is not a singular perturbation. Thus, our results should qualitatively hold also for a finite, at least not too large, asymmetry.

To confirm this finding and study intermediate values of $\rho$ (besides $\rho=0,1$ considered analytically) we performed simulations with spatial heterogeneity $\rho=0.9$ and asymmetry in the interactions $\gamma=0.9$, and as before for $L=400$, $S=200$.
Also in this case at $T=0.8$ we find a continuous transition and no strong dependence on the initial conditions, while at $T=0.18$ we find a discontinuous transition and a hysteresis region (Figure \ref{fig:numericalhasym}) \footnote{ 
At $T=0.18$ the dynamics is so slow (especially close to the tipping points) that at $t_{max}=500$ some of the abundances have not yet converged to their asymptotic values.
This leads to an apparent smoothing of the discontinuous transition, whose existence is nevertheless ensured by the abrupt change of behaviour of the evolution of the abundance (see Figure \ref{fig:hvstgamma}), analogous to the one observed for $\rho=1$, $\gamma=1$.}. 
Although the curves quantitatively change with respect to their $\gamma=\rho=1$ counterparts, as expected, the results and in particular the existence of a discontinuous transition do remain qualitatively unaltered. 

In conclusion, combining all these numerical tests, we conclude that the scenario obtained from the analytical solution is robust and holds broadly. We will come back to this point in the conclusion to suggest other extensions and tests.

\section{Precursor of the instability toward extinction}
In the previous section, we have shown that dispersal can rescue complex and large ecosystems from extinction due to demographic noise. Depending on the strength of the demographic noise, the transition from the self-sustained to the extinct phase can be either continuous or discontinuous. The latter takes place for low demographic noise and low dispersal. In this regime, we have found that the transition is accompanied by a metastable regime and hysteresis. Such transition is what is called in ecology, in environmental and social sciences a {\it tipping point} or regime-shift \cite{scheffer2001catastrophic,lenton2013environmental} and in physics a spinodal. 
Tipping points are often catastrophic events, as the abrupt rapid shifts almost always lead to negative consequences and a less favorable state of the system. Our case is no exception, as the system's transition is from a self-sustained state with high diversity to one in which all species are extinct. As done for several other tipping points \cite{scheffer2012anticipating,dakos2023tipping}, it is therefore important to find early signs or precursors that can allow one to detect the closeness of the system to the tipping point before the catastrophic shift actually takes place. 

In our case, following intuition that comes from the physics of spinodal points, we focus on responses to perturbations as probe of closeness to the tipping point. We can show analytically (see Appendix \ref{app:precursor}) that the instability of the self-sustained state is accompanied by a diverging response to perturbations. This phenomenon is strongly linked to the saddle-node bifurcation of the mean-field equations that governs the transition. 

\begin{figure}[htbp]
         \centering
         \includegraphics[width=0.45\textwidth]{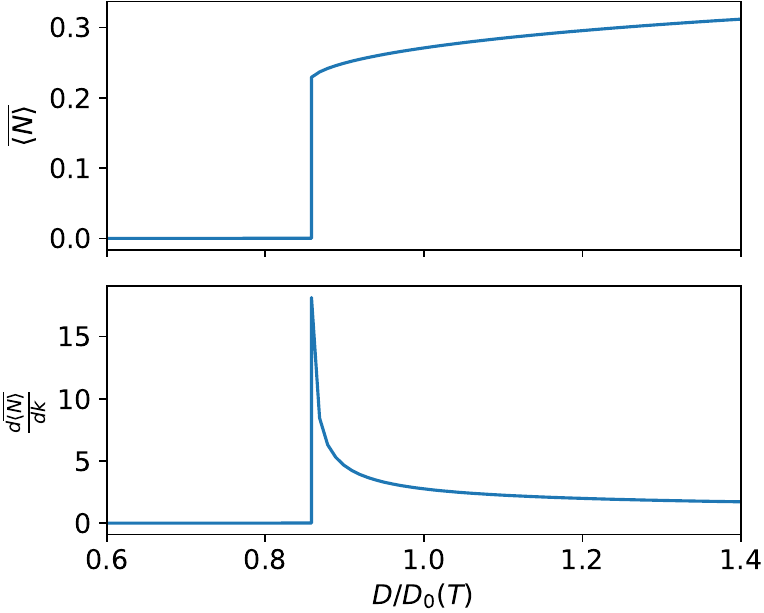}
\caption{Average abundance and its response to a perturbation of the carrying capacity $k$ at $T=0.153$ for $\rho=0$ approaching the instability of the self-sustained phase. $\mu=1$, $\sigma=0.5$.}
	\label{fig:divergentresponse}
\end{figure}

In particular, we have studied the change of the average abundance due to a change in the carrying capacity. Such response, which can be measured in controlled lab experiments, 
does 
diverge approaching the discontinuous transition, see Figure \ref{fig:divergentresponse} for the $\rho=0$ case. A similar behavior is expected for generic values of $\rho$. 
This probe can therefore be used as an early warning signal of the proximity to the tipping point of the self-sustained phase. 
In natural ecosystems, where measuring responses to perturbation can be challenging, one could instead monitor the long-term fluctuations of average abundance due to  environmental noise affecting the carrying capacity on a long time. This would be a proxy for the response proposed above (it is important to focus on long-times as all the processes at play are slow).

\section{Conclusions} 
\label{sec:conclusions}

We uncovered a rich phase diagram for many-species Lotka-Volterra metacommunities subject to heterogeneous symmetric interactions, demographic noise and diffusion. 
If the demographic fluctuations are too strong they drive all species to extinctions, but when the diffusion constant is large enough these extinctions can be compensated by recolonizations from neighboring sites, and the ecosystem is able to self-sustain at finite abundance and diversity.
The system exhibits a phase transition between an extinction and a survival phase. The transition can be either continuous or discontinuous, depending on whether the behaviour of the system is dominated by the demographic fluctuations or the heterogeneous interaction network. 

When the demographic fluctuations are strong the transition is continuous and interactions play a secondary role. In fact, the transition is completely analogous to what one would have in the absence of the interactions (even the critical value of the diffusion constant coincide).
This is because when the abundances tend to zero the interactions become sub-dominant and the system falls in the standard Directed Percolation universality class.

The situation is drastically different at lower demographic noise. In this case the transition becomes discontinuous and the system exhibits novel features, that are a signature of the complexity of the ecosystem and the major role played by the interactions. 
There is an extended range of parameters in which without interactions, i.e. for single species, the system would be driven to extinction but the metacommunity is instead able to self-sustain at finite abundances. 
This is possible because strongly competing species are eliminated from the community, while surviving species cooperate to self-sustain in such harsh conditions. 
For small demographic noise and lowering the diffusion constant, the ecosystem reaches a tipping point at which all surviving species go extinct; close to this point the ecosystem is subject to collapses upon small perturbations and its dynamics exhibits hysteresis. 
We therefore find that mutualism naturally emerges from a (on average) competitive pool of species when conditions become harsher. 
This has a double effect: it allows the ecosystem to survive in conditions in which all species in isolation would go extinct, but it also makes it fragile to perturbations. In this regime, it is not possible to predict the vicinity of the catastrophic shift of the ecosystem by looking at the average abundance. As early warning sign, we propose to monitor the response of the system to perturbations. We have shown that this is a suitable probe, as it diverges approaching the discontinuous transition.


We confirm and complement our analytical approach with numerical simulations, which show 
that our results are quite robust to modifications of the model, in particular to the introduction of a small asymmetry in the interactions, to various degrees of correlation of the interaction network between different spatial locations, and for system with a finite number of species and patches.   

There are several directions worth future investigations. We focused on a fully connected spatial system, which provides a  mean-field analysis for generic spatial lattices. On the other hand, our DMFT treatment of the interactions is directly generalizable to any other spatial network, including finite dimensional ones. It would be very interesting to study cases in which the patches are located in finite dimensional lattice or on random structures. In particular, it would be interesting to find out (1)
whether the discontinuous transition is also present in this case or finite dimensional fluctuations destroy the metastable region, and (2) whether the continuous transition can still be described in terms of  directed percolation or interactions, although secondary, can alter its universality class. It would also be worth analysing stronger asymmetries in the interactions, e.g. lowering the value of $\gamma$. We expect that a significant positive correlation between reciprocal interactions is needed to induce metastability. This ensures that species that interact more competitively are also more negatively affected by the interactions with the rest of the ecosystem and hence go extinct, thus leading to mutualism for the surviving species.

Finally, species rich LV model with heterogeneous and strong interactions display multiple equilibria and chaotic dynamics \cite{Kessler2015, bunin2017, biroli2018, altieri2021}.
The possibility of different patches to converge to different stationary states could strongly modify the behaviour of the system; in particular allowing the system to experience higher values of the global diversity, possibly violating May's bound \cite{May1972}.

\vspace{0.5cm}

\section*{Acknowledgments} 
During the preparation of the manuscript we became aware of Jonas Denk and Oskar Hallatscheck work on tipping points in mutualistic Lotka-Volterra communities \cite{denk2023}. 
Their results are complementary and agree with ours. 
We would like to thank them for sharing their results and constructive interactions. 
We also thank Joseph Baron, M. Barbier, J. F. Arnoldi, and L. F. Cugliandolo for stimulating discussions.

This work was supported by the Simons Foundation Grant on \emph{Cracking the Glass Problem} (\# 454935 Giulio Biroli). 

\bibliographystyle{unsrt}
\bibliography{MyLibrary}

\begin{thebibliography}{10}

\bibitem{lozupone2012}
Catherine~A. Lozupone, Jesse~I. Stombaugh, Jeffrey~I. Gordon, Janet~K. Jansson,
  and Rob Knight.
\newblock Diversity, stability and resilience of the human gut microbiota.
\newblock {\em Nature}, 489(7415):220--230, September 2012.

\bibitem{Kessler2015}
David~A Kessler and Nadav~M Shnerb.
\newblock Generalized model of island biodiversity.
\newblock {\em Physical Review E}, 91(4):042705, April 2015.

\bibitem{bunin2017}
Guy Bunin.
\newblock Ecological communities with {{Lotka-Volterra}} dynamics.
\newblock {\em Physical Review E}, 95(4):1--8, 2017.

\bibitem{biroli2018}
Giulio Biroli, Guy Bunin, and Chiara Cammarota.
\newblock Marginally stable equilibria in critical ecosystems.
\newblock {\em New Journal of Physics}, 20(8):083051, August 2018.

\bibitem{altieri2021}
Ada Altieri, Felix Roy, Chiara Cammarota, and Giulio Biroli.
\newblock Properties of {{Equilibria}} and {{Glassy Phases}} of the {{Random
  Lotka-Volterra Model}} with {{Demographic Noise}}.
\newblock {\em Physical Review Letters}, 126(25):258301, June 2021.

\bibitem{galla2018}
Tobias Galla.
\newblock Dynamically evolved community size and stability of random
  {{Lotka-Volterra}} ecosystems(a).
\newblock {\em Epl}, 123(4):1--13, 2018.

\bibitem{rogers2022}
Tanya~L. Rogers, Bethany~J. Johnson, and Stephan~B. Munch.
\newblock Chaos is not rare in natural ecosystems.
\newblock {\em Nature Ecology \& Evolution}, 6(8):1105--1111, August 2022.

\bibitem{gross2005}
Thilo Gross, Wolfgang Ebenh{\"o}h, and Ulrike Feudel.
\newblock Long food chains are in general chaotic.
\newblock {\em Oikos}, 109(1):135--144, 2005.

\bibitem{leibold2004}
M.~A. Leibold, M.~Holyoak, N.~Mouquet, P.~Amarasekare, J.~M. Chase, M.~F.
  Hoopes, R.~D. Holt, J.~B. Shurin, R.~Law, D.~Tilman, M.~Loreau, and
  A.~Gonzalez.
\newblock The metacommunity concept: A framework for multi-scale community
  ecology.
\newblock {\em Ecology Letters}, 7(7):601--613, 2004.

\bibitem{hassell1991}
Michael~P. Hassell, Hugh~N. Comins, and Robert~M. Mayt.
\newblock Spatial structure and chaos in insect population dynamics.
\newblock {\em Nature}, 353(6341):255--258, September 1991.

\bibitem{mobilia2007}
Mauro Mobilia, Ivan~T Georgiev, and Uwe~C T{\"a}uber.
\newblock Phase {{Transitions}} and {{Spatio-Temporal Fluctuations}} in
  {{Stochastic Lattice Lotka}}\textendash{{Volterra Models}}.
\newblock {\em Journal of Statistical Physics}, 128(1-2):447--483, June 2007.

\bibitem{olmeda2023}
Fabrizio Olmeda and Steffen Rulands.
\newblock Long-range interactions and disorder facilitate pattern formation in
  spatial complex systems, March 2023.

\bibitem{dobramysl2018}
Ulrich Dobramysl, Mauro Mobilia, Michel Pleimling, and Uwe~C. Tauber.
\newblock Stochastic population dynamics in spatially extended predator-prey
  systems.
\newblock {\em Journal of Physics A: Mathematical and Theoretical}, 51(6),
  2018.

\bibitem{Roy2020}
Felix Roy, Matthieu Barbier, Giulio Biroli, and Guy Bunin.
\newblock Complex interactions can create persistent fluctuations in
  high-diversity ecosystems.
\newblock {\em PLoS Computational Biology}, 16(5):1--14, 2020.

\bibitem{pearce2020}
Michael~T. Pearce, Atish Agarwala, and Daniel~S. Fisher.
\newblock Stabilization of extensive fine-scale diversity by ecologically
  driven spatiotemporal chaos.
\newblock {\em Proceedings of the National Academy of Sciences},
  117(25):14572--14583, June 2020.

\bibitem{denk2022}
Jonas Denk and Oskar Hallatschek.
\newblock Self-consistent dispersal puts tight constraints on the
  spatiotemporal organization of species-rich metacommunities.
\newblock {\em Proceedings of the National Academy of Sciences},
  119(26):e2200390119, June 2022.

\bibitem{baron2020}
Joseph~W. Baron and Tobias Galla.
\newblock Dispersal-induced instability in complex ecosystems.
\newblock {\em Nature Communications}, 11(1):6032, November 2020.

\bibitem{leemput2015}
Ingrid~A. van~de Leemput, Egbert~H. van Nes, and Marten Scheffer.
\newblock Resilience of {{Alternative States}} in {{Spatially Extended
  Ecosystems}}.
\newblock {\em PLOS ONE}, 10(2):e0116859, February 2015.

\bibitem{may1974}
Robert~M. May.
\newblock {\em Stability and {{Complexity}} in {{Model Ecosystems}}}, volume~1.
\newblock {Princeton University Press}, 1974.

\bibitem{grilli2020}
Jacopo Grilli.
\newblock Macroecological laws describe variation and diversity in microbial
  communities.
\newblock {\em Nature Communications}, 11(1):4743, December 2020.

\bibitem{azaele2016}
Sandro Azaele, Samir Suweis, Jacopo Grilli, Igor Volkov, Jayanth~R. Banavar,
  and Amos Maritan.
\newblock Statistical mechanics of ecological systems: {{Neutral}} theory and
  beyond.
\newblock {\em Reviews of Modern Physics}, 88(3):035003, July 2016.

\bibitem{kamenev2008}
Alex Kamenev, Baruch Meerson, and Boris Shklovskii.
\newblock How {{Colored Environmental Noise Affects Population Extinction}}.
\newblock {\em Physical Review Letters}, 101(26):268103, December 2008.

\bibitem{larroya2023}
Ferran Larroya and Tobias Galla.
\newblock Demographic noise in complex ecological communities.
\newblock {\em Journal of Physics: Complexity}, 2023.

\bibitem{vasseur2004}
David~A. Vasseur and Peter Yodzis.
\newblock The {{Color}} of {{Environmental Noise}}.
\newblock {\em Ecology}, 85(4):1146--1152, 2004.

\bibitem{petchey1997}
O.~L. Petchey, A.~Gonzalez, and H.~B. Wilson.
\newblock Effects on population persistence: The interaction between
  environmental noise colour, intraspecific competition and space.
\newblock {\em Proceedings of the Royal Society B: Biological Sciences},
  264(1389):1841--1847, December 1997.

\bibitem{realpe-gomez2013}
John {Realpe-Gomez}, Mara Baudena, Tobias Galla, Alan~J. McKane, and Max
  Rietkerk.
\newblock Demographic noise and resilience in a semi-arid ecosystem model.
\newblock {\em Ecological Complexity}, 15:97--108, September 2013.

\bibitem{bell2000}
Graham Bell and Associate Editor:~Dolph Schluter.
\newblock The {{Distribution}} of {{Abundance}} in {{Neutral Communities}}.
\newblock {\em The American Naturalist}, 155(5):606--617, 2000.

\bibitem{shoemaker2020}
Lauren~G. Shoemaker, Lauren~L. Sullivan, Ian Donohue, Juliano~S. Cabral,
  Ryan~J. Williams, Margaret~M. Mayfield, Jonathan~M. Chase, Chengjin Chu,
  W.~Stanley Harpole, Andreas Huth, Janneke HilleRisLambers, Aubrie R.~M.
  James, Nathan J.~B. Kraft, Felix May, Ranjan Muthukrishnan, Sean Satterlee,
  Franziska Taubert, Xugao Wang, Thorsten Wiegand, Qiang Yang, and Karen~C.
  Abbott.
\newblock Integrating the underlying structure of stochasticity into community
  ecology.
\newblock {\em Ecology}, 101(2):e02922, 2020.

\bibitem{peruzzo2020}
Fabio Peruzzo, Mauro Mobilia, and Sandro Azaele.
\newblock Spatial {{Patterns Emerging}} from a {{Stochastic Process Near
  Criticality}}.
\newblock {\em Physical Review X}, 10(1):011032, February 2020.

\bibitem{macarthur1967}
Robert~H. MacArthur and Edward~O. Wilson.
\newblock {\em The {{Theory}} of {{Island Biogeography}}}.
\newblock {Princeton University Press}, 1967.

\bibitem{hu2022}
Jiliang Hu, Daniel~R. Amor, Matthieu Barbier, Guy Bunin, and Jeff Gore.
\newblock Emergent phases of ecological diversity and dynamics mapped in
  microcosms.
\newblock {\em Science}, 378(6615):85--89, October 2022.

\bibitem{garcialorenzana2022}
Giulia Garcia~Lorenzana and Ada Altieri.
\newblock Well-mixed {{Lotka-Volterra}} model with random strongly competitive
  interactions.
\newblock {\em Physical Review E}, 105(2):024307, February 2022.

\bibitem{loreau_biodiversity_2003}
Michel Loreau, Nicolas Mouquet, and Andrew Gonzalez.
\newblock Biodiversity as spatial insurance in heterogeneous landscapes.
\newblock {\em Proceedings of the National Academy of Sciences},
  100(22):12765--12770, 2003.
\newblock Publisher: Proceedings of the National Academy of Sciences.

\bibitem{chesson_general_2000}
Peter Chesson.
\newblock General theory of competitive coexistence in spatially-varying
  environments.
\newblock {\em Theoretical Population Biology}, 58(3):211--237, 2000.

\bibitem{gravel2016}
Dominique Gravel, Fran{\c c}ois Massol, and Mathew~A. Leibold.
\newblock Stability and complexity in model meta-ecosystems.
\newblock {\em Nature Communications}, 7(1):12457, November 2016.

\bibitem{pettersson2021}
Susanne Pettersson and Martin~Nilsson Jacobi.
\newblock Spatial heterogeneity enhance robustness of large multi-species
  ecosystems.
\newblock {\em PLOS Computational Biology}, 17(10):e1008899, October 2021.

\bibitem{mahadevan_spatiotemporal_2023}
Aditya Mahadevan, Michael~T Pearce, and Daniel~S Fisher.
\newblock Spatiotemporal ecological chaos enables gradual evolutionary
  diversification without niches or tradeoffs.
\newblock {\em eLife}, 12:e82734, April 2023.
\newblock Publisher: eLife Sciences Publications, Ltd.

\bibitem{broadbent1957}
S.~R. Broadbent and J.~M. Hammersley.
\newblock Percolation processes: {{I}}. {{Crystals}} and mazes.
\newblock {\em Mathematical Proceedings of the Cambridge Philosophical
  Society}, 53(3):629--641, July 1957.

\bibitem{janssen1997}
H.~K. Janssen.
\newblock Spontaneous {{Symmetry Breaking}} in {{Directed Percolation}} with
  {{Many Colors}}: {{Differentiation}} of {{Species}} in the {{Gribov
  Process}}.
\newblock {\em Physical Review Letters}, 78(15):2890--2893, April 1997.

\bibitem{hinrichsen2000}
Haye Hinrichsen.
\newblock Non-equilibrium critical phenomena and phase transitions into
  absorbing states.
\newblock {\em Advances in Physics}, 49(7):815--958, November 2000.

\bibitem{scheffer2001}
Marten Scheffer, Steve Carpenter, Jonathan~A. Foley, Carl Folke, and Brian
  Walker.
\newblock Catastrophic shifts in ecosystems.
\newblock {\em Nature}, 413(6856):591--596, October 2001.

\bibitem{kefi2007}
Sonia K{\'e}fi, Max Rietkerk, Minus {van Baalen}, and Michel Loreau.
\newblock Local facilitation, bistability and transitions in arid ecosystems.
\newblock {\em Theoretical Population Biology}, 71(3):367--379, May 2007.

\bibitem{lenton2008}
Timothy~M. Lenton, Hermann Held, Elmar Kriegler, Jim~W. Hall, Wolfgang Lucht,
  Stefan Rahmstorf, and Hans~Joachim Schellnhuber.
\newblock Tipping elements in the {{Earth}}'s climate system.
\newblock {\em Proceedings of the National Academy of Sciences},
  105(6):1786--1793, February 2008.

\bibitem{bouchaud2013}
Jean-Philippe Bouchaud.
\newblock Crises and {{Collective Socio-Economic Phenomena}}: {{Simple Models}}
  and {{Challenges}}.
\newblock {\em Journal of Statistical Physics}, 151(3):567--606, May 2013.

\bibitem{denk2023}
Jonas Denk and Oskar Hallatschek.
\newblock Tipping points emerge from weak mutualism in metacommunities.
\newblock Preprint, {Ecology}, February 2023.

\bibitem{mezard1986}
M~Mezard, G~Parisi, and M~Virasoro.
\newblock {\em Spin {{Glass Theory}} and {{Beyond}}: {{An Introduction}} to the
  {{Replica Method}} and {{Its Applications}}}, volume~9 of {\em World
  {{Scientific Lecture Notes}} in {{Physics}}}.
\newblock {WORLD SCIENTIFIC}, November 1986.

\bibitem{May1972}
Robert~M. May.
\newblock Will a {{Large Complex System}} be {{Stable}}?
\newblock {\em Nature}, 238(5364):413--414, August 1972.

\bibitem{altieri2022}
Ada Altieri and Giulio Biroli.
\newblock Effects of intraspecific cooperative interactions in large
  ecosystems.
\newblock {\em SciPost Physics}, 12(1):013, January 2022.

\bibitem{fisher2014}
Charles~K. Fisher and Pankaj Mehta.
\newblock The transition between the niche and neutral regimes in ecology.
\newblock {\em Proceedings of the National Academy of Sciences of the United
  States of America}, 111(36):13111--13116, 2014.

\bibitem{pigani2022}
Emanuele Pigani, Damiano Sgarbossa, Samir Suweis, Amos Maritan, and Sandro
  Azaele.
\newblock Delay effects on the stability of large ecosystems.
\newblock {\em Proceedings of the National Academy of Sciences},
  119(45):e2211449119, November 2022.

\bibitem{suweis2023}
Samir Suweis, Francesco Ferraro, Sandro Azaele, and Amos Maritan.
\newblock Generalized {{Lotka-Volterra Systems}} with {{Time Correlated
  Stochastic Interactions}}, July 2023.

\bibitem{barbier2018}
Matthieu Barbier, Jean~Fran{\c c}ois Arnoldi, Guy Bunin, and Michel Loreau.
\newblock Generic assembly patterns in complex ecological communities.
\newblock {\em Proceedings of the National Academy of Sciences of the United
  States of America}, 115(9):2156--2161, 2018.

\bibitem{Behn1992}
Ulrich Behn, J.~Leo {van Hemmen}, and Bernhard Sulzer.
\newblock Memory {{B Cells Stabilize Cycles}} in a {{Repressive Network}}.
\newblock In {\em Theoretical and {{Experimental Insights}} into
  {{Immunology}}}, pages 249--260. {Springer Berlin Heidelberg}, {Berlin,
  Heidelberg}, 1992.

\bibitem{moran2019}
Jos{\'e} Moran and Jean-Philippe Bouchaud.
\newblock May's instability in large economies.
\newblock {\em Physical Review E}, 100(3):032307, September 2019.

\bibitem{Goodwin1990}
Richard~M. Goodwin.
\newblock {\em Chaotic {{Economic Dynamics}}}.
\newblock {Oxford University Press}, November 1990.

\bibitem{Bomze1995}
Immanuel~M. Bomze.
\newblock Lotka-{{Volterra}} equation and replicator dynamics: New issues in
  classification.
\newblock {\em Biological Cybernetics}, 72(5):447--453, 1995.

\bibitem{bunin2016}
Guy Bunin.
\newblock Interaction patterns and diversity in assembled ecological
  communities, July 2016.

\bibitem{feller1951}
William Feller.
\newblock Two {{Singular Diffusion Problems}}.
\newblock {\em Annals of Mathematics}, 54(1):173--182, 1951.

\bibitem{dornic2005}
Ivan Dornic, Hugues Chat{\'e}, and Miguel~A. Mu{\~n}oz.
\newblock Integration of {{Langevin Equations}} with {{Multiplicative Noise}}
  and the {{Viability}} of {{Field Theories}} for {{Absorbing Phase
  Transitions}}.
\newblock {\em Physical Review Letters}, 94(10):100601, March 2005.

\bibitem{cardy1980}
J~L Cardy and R~L Sugar.
\newblock Directed percolation and {{Reggeon}} field theory.
\newblock {\em Journal of Physics A: Mathematical and General},
  13(12):L423--L427, December 1980.

\bibitem{roy2019}
F~Roy, G~Biroli, G~Bunin, and C~Cammarota.
\newblock Numerical implementation of dynamical mean field theory for
  disordered systems: Application to the {{Lotka}}\textendash{{Volterra}} model
  of ecosystems.
\newblock {\em Journal of Physics A: Mathematical and Theoretical},
  52(48):484001, November 2019.

\bibitem{zwanzig2001}
Robert Zwanzig.
\newblock {\em Nonequilibrium Statistical Mechanics}.
\newblock {Oxford Univ. Press}, {Oxford}, 2001.

\bibitem{georges1996}
Antoine Georges, Gabriel Kotliar, Werner Krauth, and Marcelo~J. Rozenberg.
\newblock Dynamical mean-field theory of strongly correlated fermion systems
  and the limit of infinite dimensions.
\newblock {\em Reviews of Modern Physics}, 68(1):13--125, January 1996.

\bibitem{cugliandolo2023}
Leticia~F. Cugliandolo.
\newblock Recent {{Applications}} of {{Dynamical Mean-Field Methods}}, May
  2023.

\bibitem{arous1997symmetric}
G~Ben Arous and Alice Guionnet.
\newblock Symmetric langevin spin glass dynamics.
\newblock {\em The Annals of Probability}, 25(3):1367--1422, 1997.

\bibitem{chen2016}
Sheng Chen and Uwe~C. T{\"a}uber.
\newblock Non-equilibrium relaxation in a stochastic lattice
  {{Lotka}}\textendash{{Volterra}} model.
\newblock {\em Physical Biology}, 13(2):025005, April 2016.

\bibitem{martin1973}
P.~C. Martin, E.~D. Siggia, and H.~A. Rose.
\newblock Statistical {{Dynamics}} of {{Classical Systems}}.
\newblock {\em Physical Review A}, 8(1):423--437, July 1973.

\bibitem{janssen1976}
Hans-Karl Janssen.
\newblock On a {{Lagrangean}} for classical field dynamics and renormalization
  group calculations of dynamical critical properties.
\newblock {\em Zeitschrift f\"ur Physik B Condensed Matter}, 23(4):377--380,
  December 1976.

\bibitem{dedominicis1978}
C.~De~Dominicis.
\newblock Dynamics as a substitute for replicas in systems with quenched random
  impurities.
\newblock {\em Physical Review B}, 18(9):4913--4919, November 1978.

\bibitem{aron2010}
Camille Aron, Giulio Biroli, and Leticia~F. Cugliandolo.
\newblock Symmetries of generating functionals of {{Langevin}} processes with
  colored multiplicative noise.
\newblock {\em Journal of Statistical Mechanics: Theory and Experiment},
  2010(11):P11018, November 2010.

\bibitem{wu_understanding_2021}
Jim Wu, Pankaj Mehta, and David Schwab.
\newblock Understanding {Species} {Abundance} {Distributions} in {Complex}
  {Ecosystems} of {Interacting} {Species}, March 2021.
\newblock arXiv:2103.02081 [q-bio].

\bibitem{baron2023}
Joseph~W. Baron, Thomas~Jun Jewell, Christopher Ryder, and Tobias Galla.
\newblock Breakdown of {{Random-Matrix Universality}} in {{Persistent
  Lotka-Volterra Communities}}.
\newblock {\em Physical Review Letters}, 130(13):137401, March 2023.

\bibitem{Note1}
The discontinuous transition takes place slightly before the analytical
  prediction. Besides finite size effects, we note that this phenomenon is to
  be expected for this kind of transition. In fact, when the red curve (low
  initial condition) jumps to high abundance, this does not necessarily
  indicate that the $\protect \overline {\langle N\rangle }=0$ solution has
  become locally unstable, but rather that its basin of attraction has shrunk
  and does not include the considered initial condition anymore. It is
  therefore to be expected that this occurs for $D<D_0(T)$. A similar
  phenomenon takes place for spinodal transition in physics.

\bibitem{Note2}
At $T=0.18$ the dynamics is so slow (especially close to the tipping points)
  that at $t_{max}=500$ some of the abundances have not yet converged to their
  asymptotic values. This leads to an apparent smoothing of the discontinuous
  transition, whose existence is nevertheless ensured by the abrupt change of
  behaviour of the evolution of the abundance (see Figure \ref
  {fig:hvstgamma}), analogous to the one observed for $\rho =1$, $\gamma =1$.

\bibitem{scheffer2001catastrophic}
Marten Scheffer, Steve Carpenter, Jonathan~A Foley, Carl Folke, and Brian
  Walker.
\newblock Catastrophic shifts in ecosystems.
\newblock {\em Nature}, 413(6856):591--596, 2001.

\bibitem{lenton2013environmental}
Timothy~M Lenton.
\newblock Environmental tipping points.
\newblock {\em Annual Review of Environment and Resources}, 38:1--29, 2013.

\bibitem{scheffer2012anticipating}
Marten Scheffer, Stephen~R Carpenter, Timothy~M Lenton, Jordi Bascompte,
  William Brock, Vasilis Dakos, Johan Van~de Koppel, Ingrid~A Van~de Leemput,
  Simon~A Levin, Egbert~H Van~Nes, et~al.
\newblock Anticipating critical transitions.
\newblock {\em science}, 338(6105):344--348, 2012.

\bibitem{dakos2023tipping}
Vasilis Dakos, Chris~A Boulton, Josh~E Buxton, Jesse~F Abrams, David~I
  Armstrong~McKay, Sebastian Bathiany, Lana Blaschke, Niklas Boers, Daniel
  Dylewsky, Carlos L{\'o}pez-Mart{\'\i}nez, et~al.
\newblock Tipping point detection and early-warnings in climate, ecological,
  and human systems.
\newblock {\em EGUsphere}, 2023:1--35, 2023.

\bibitem{altieri2020}
Ada Altieri, Giulio Biroli, and Chiara Cammarota.
\newblock Dynamical mean-field theory and aging dynamics.
\newblock {\em Journal of Physics A: Mathematical and Theoretical},
  53(37):375006, September 2020.

\bibitem{aron2016}
Camille Aron, Daniel~G. Barci, Leticia~F. Cugliandolo, Zochil~Gonz{\'a}lez
  Arenas, and Gustavo~S. Lozano.
\newblock Dynamical symmetries of {{Markov}} processes with multiplicative
  white noise.
\newblock {\em Journal of Statistical Mechanics: Theory and Experiment},
  2016(5):053207, May 2016.

\bibitem{weissmann2018}
Haim Weissmann, Nadav~M. Shnerb, and David~A. Kessler.
\newblock Simulation of spatial systems with demographic noise.
\newblock {\em Physical Review E}, 98(2):022131, August 2018.

\end{thebibliography}

\appendix
 \begin{widetext}

\section{DMFT derivation}

\label{app:DMFT}
Here we outline the derivation, adapted from reference \cite{roy2019}, of the Dynamical Mean Field Theory for our system, for generic value of the spatial correlation of the interactions $\rho$.

We consider $S$ species, indexed by $i=1,...S$, and their Lotka-Volterra dynamics,
    \begin{align}
    \dot{N}_{i,u}=N_{i,u}\left(1-N_{i,u}-\sum_j\alpha_{ij}^u N_{j,u} +\zeta_{i,u}\right)+ D\left(\frac{1}{c}\sum_{v\in \partial u} N_{i,v}-N_{i,u}\right) + \eta_i^u(t)\sqrt{N_{i,u}} + \lambda_{i,u}        
      \end{align}
to which we have added a perturbation to the carrying capacity $\zeta_{i,u}$ and an external immigration $\lambda_{i,u}$, that will be taken to zero at the end of the computation.
These equations (for a given value of the $\eta_i^u(t)$) define the trajectories $N_{i,u}(t)$.
We add a new species, $i=0$, to the system, and we draw its interactions and initial conditions independently from the rest of the system and with the same statistics. 
At large $S$, thanks to the scaling of the interactions, the presence of a new species is a small perturbation to the system, so that the trajectories of the other $S$ species will only be slightly modified. 
We consider their linear response:
\begin{align}
    \delta N_{i,u}(t)=\sum_{v\in \partial u, j} \int_0^t\frac{\delta N_{i,u}(t)}{\delta \zeta_{j,v}(t')}(-\alpha_{j0}^v N_{0,v}(t'))dt'=\sum_{v\in \partial u, i} \int_0^tR_{i,j}^{u,v}(t,t')(-\alpha_{j0}^v N_{0,v}(t'))dt'
\end{align}
We have introduced the response function $R_{i,j}^{u,v}(t,t')$ of the abundance of species $i$ in patch $u$ at time $t$ to a variation in the carrying capacity of species $j$ in patch $v$ at time $t'$.

The dynamics of species $0$ will depend on these new trajectories:
\begin{align}
    \dot{N}_{0,u}=N_{0,u}\left(1-N_{0,u}-\sum_i\alpha_{0i}^u \left(N^0_{i,u}+\delta N_{i,u}\right)\right) + D\left(\frac{1}{c}\sum_{v\in \partial u} N_{0,v}-N_{0,u}\right) + \eta_{0,u}(t)\sqrt{N_{0,u}} \ .
\end{align}

Because the correlations between interaction coefficients in any two patches are the same, these Gaussian variables can generically be decomposed into a common random contribution, identical in all patches and proportional to the correlation $\rho$, and one independent in different patches, proportional to $\sqrt{1-\rho^2}$.
We thus introduce the matrix $a_{i,j}$ and $a_{i,j}^u$ such that $\alpha_{i,j}^u=\mu/S+\sigma \left(\rho a_{i,j}+\sqrt{1-\rho^2} a_{i,j}^u\right)$, $\mathbb{E}\left[a_{i,j}\right]=\mathbb{E}\left[a_{i,j}^u\right]=0$, $\mathbb{E}\left[a_{i,j}^2\right]=\mathbb{E}\left[{a_{i,j}^u}^2\right]=1/S$, $\mathbb{E}\left[a_{i,j}a_{j,i}\right]=\mathbb{E}\left[a_{i,j}^ua_{j,i}^u\right]=\gamma/S$ and all other correlations are 0.
We can rewrite the interaction term as:
\begin{align}
\label{eq:interterm}
\begin{split}
    -\sum_i\alpha_{0i} \left(N^0_{i,u}+\delta N_{i,u}\right)=-\sum_i\left(\mu/S + \sigma  \left(\rho a_{0i}+\sqrt{1-\rho^2} a_{0i}^u\right)\right)N^0_{i,u}+\\+\sum_{i,j}\left(\mu/S + \sigma  \left(\rho a_{i0}+\sqrt{1-\rho^2} a_{i0}^u\right)\right)\left(\mu/S + \sigma  \left(\rho a_{0j}+\sqrt{1-\rho^2} a_{0j}^u\right)\right)\sum_{v\in \partial u} \int_0^tR_{i,j}^{u,v}(t,t') N_{0,v}(t')dt' \ .
\end{split} 
\end{align}

We want to describe its statistical properties in the limit $S\to\infty$. 
The response function $R_{i,j}^{u,v}(t,t')$ is defined on the unperturbed trajectories, and is therefore uncorrelated from the interactions coefficients with species $0$. 
$R_{i,j}^{u,v}(t,t')\sim 1/\sqrt{S}$ for $i\neq j$ \cite{roy2019}, so that the off-diagonal terms can be neglected.
Thanks to the central limit theorem, $\sum_ja_{0j} a_{j0}R_{j,j}^{u,v}(t,t')$ will converge to its average:
\begin{align}
    \sum_j a_{0j} a_{j0}R_{j,j}^{u,v}(t,t')\to S \mathbb{E}\left[ a_{0j} a_{j0}\right]\mathbb{E}\left[ R_{j,j}^{u,v}(t,t')\right]=\gamma\mathbb{E}\left[ R_{j,j}^{u,v}(t,t')\right] \ .
\end{align}
By similarly evaluating all terms in (\ref{eq:interterm}) we obtain:
\begin{align}
\label{eq:intDMFT}
\begin{split}
    -\sum_j\alpha_{0j} \left(N^0_{j,u}+\delta N_{j,u}\right)\to -\mu\mathbb{E}\left[ N^0_{j,u}\right] - \sigma \rho \tilde{\xi}_u(t) - \sigma \sqrt{1-\rho^2} \tilde{\psi}_u(t) +\\ +\sigma ^2\rho^2\gamma\sum_{v\in \partial u} \int_0^t\mathbb{E}\left[R_{j,j}^{u,v}(t,t')\right] N_{0,v}(t')dt'  +\sigma ^2(1-\rho^2)\gamma\int_0^t\mathbb{E}\left[R_{j,j}^{u,u}(t,t')\right] N_{0,u}(t')dt' \ , 
\end{split}
\end{align}
where $\tilde{\xi}_u(t)$ and $\tilde{\psi}_u(t)$ are Gaussian fields with zero mean and covariance $\mathbb{E}\left[\tilde{\xi}_u(t)\tilde{\xi}_v(t')\right]=\mathbb{E}\left[N^0_{j,u}(t)N^0_{j,v}(t')\right]$, $\mathbb{E}\left[\tilde{\psi}_u(t)\tilde{\psi}_v(t')\right]=\delta_{uv}\mathbb{E}\left[N^0_{j,u}(t)N^0_{j,u}(t')\right]$. 
Note that $\tilde{\xi}_u$ and the first integral of \ref{eq:intDMFT} derive from the component of the interactions constant across patches, $a_{ij}$, as we can see from the $\rho$-dependent prefactors, and they therefore couple different patches. $\tilde{\psi}_u$ and the second integral of \ref{eq:intDMFT} derive instead from the component of the interactions independent across patches, $a_{ij}^u$, and therefore represent diagonal correlations and responses.
Plugging this expression in the dynamical equation for species $0$ we obtain:
\begin{align}
\begin{split}
    \dot{N}_{0,u}=N_{0,u}\Bigg(1-N_{0,u}-\mu\mathbb{E}\left[ N^0_{j,u}\right] - \sigma \rho \tilde{\xi}_u(t) - \sigma \sqrt{1-\rho^2} \tilde{\psi}_u(t) +\\ +\sigma ^2\rho^2\gamma\sum_{v\in \partial u} \int_0^t\mathbb{E}\left[R_{j,j}^{u,v}(t,t')\right] N_{0,v}(t')dt'  +\sigma ^2(1-\rho^2)\gamma\int_0^t\mathbb{E}\left[R_{j,j}^{u,u}(t,t')\right] N_{0,u}(t')dt'\Bigg) +\\+ D\left(\frac{1}{c}\sum_{v\in \partial u} N_{0,v}-N_{0,u}\right) + \eta_{0,u}(t)\sqrt{N_{0,u}}    \ . 
\end{split}
\end{align}

Species $0$ is statistically equivalent to all the others, we can therefore replace the averages over the $S$ original species with averages with respect to this new dynamics for a single species, obtaining some self-consistent equations:
\begin{align}
    \dot{N_u}=& N_u\Bigg(1-N_u-\mu h_u-\sigma \rho \tilde{\xi}_u(t) - \sigma \sqrt{1-\rho^2} \tilde{\psi}_u(t) + \sigma^2\rho^2 \gamma\int_0^t \sum_{v\in \partial u} R_{uv}(t,s)N_v(s)ds+ \\&+\sigma^2(1-\rho^2) \gamma\int_0^t R_{uu}(t,s)N_u(s)ds\Bigg)+ D \left(\frac{1}{c}\sum_{v\in \partial u} N_v -N_u\right) + \eta_u(t)\sqrt{N_u}\\
    \langle \tilde{\xi}_u(t)\tilde{\xi}_v(s)\rangle=&C_{uv}(t-s)=\mathbb{E}[N_u(t)N_v(s)]\\
    \langle \tilde{\psi}_u(t)\tilde{\psi}_v(s)\rangle=&\delta_{uv}C_{uu}(t-s)\\
    R_{uv}(t,s)=&\mathbb{E}\left[\frac{\delta N_u(t)}{\delta \zeta_v(s)}\bigg|_{\zeta=0}\right]\\
    h_u=&\mathbb{E}[N_u] \ .
\end{align}
Since species have been effectively decoupled, we can suppress the species index.

In the single equilibrium phase, we expect the process to reach a time translation invariant regime, in which the one-time averages are time-independent and two-times observables only depend on the times difference. This was shown in \mbox{\cite{altieri2021}} for a single community with demographic noise and fixed immigration and it is known to be the case for Directed Percolation \mbox{\cite{hinrichsen2000}} and in a many-species metacommunity with constant interactions \cite{denk2022}.
It is also confirmed by our numerical results that show a quick relaxation of one-time observables to an asymptotic value (see Appendix \mbox{\ref{app:numres}}), at least away from phase transitions.
Since the auto-correlation of the abundance of one species doesn't tend to zero at large times, we can decompose $\xi_u$ and $\psi_u$ into a constant and a fluctuating component:
\begin{align}
    \tilde{\xi}_u(t)= \hat{\xi}_u+\xi_u(t)\\
    \tilde{\psi}_u(t)= \hat{\psi}_u+\psi_u(t) \ ,
\end{align}
where $ \hat{\xi}_u$ and  $ \hat{\psi}_u$ are (time independent) Gaussian variables with zero mean and correlations $\lim_{\tau\to\infty}C_{uv}(t, t+\tau)=C_{uv}^\infty$ and $\delta_{uv}C_{uu}^\infty$ and the auto-correlation of $\xi_u$ and $\psi_u$ go to zero at long times. 
Averaging over $\xi_u$, $\psi_u$ and $\eta$ at fixed $\hat{\xi}_u$ and  $\hat{\psi}_u$ corresponds to performing a time-average for one species in one patch, averaging also over $\hat{\psi}_u$ and $\hat{\xi}_u$ corresponds to averaging over patches and species. In this sense $ \hat{\xi}_u$ and  $ \hat{\psi}_u$ play the role of the quenched disorder, that was previously represented by the interaction matrix $\alpha_{ij}^u$. We will refer to the average over $\xi$ and $\eta$ at fixed $\hat{\xi}_u$ and  $\hat{\psi}_u$ as \textit{thermal average} and indicate it with brackets, and to the average over $\hat{\xi}_u$ and  $\hat{\psi}_u$ as \textit{disorder average} and indicate it with an overline.

While the derivation is so far valid for any spatial network, we will now restricted ourselves to a fully connected network, in which the empirical average over neighbors can be replaced by its thermal average.
In the large $L$ limit the connected correlation over thermal fluctuations between $N_u$ and $N_v$ is sub-dominant, so that $\xi_u$ and $\xi_v$ become independent. 
A perturbation in patch $v$ influences the abundance in patch $u$ through the diffusion term, that in a fully connected network is of order $1/L$, therefore $R_{uv}$ for $u\neq v$ scales as $1/L$, whereas $R_{uu}$ is of order 1.
Since all patches are equivalent, the elements of the $R_{uv}$ matrix can only take two values:
\begin{align}
    R_{uu}&= R_d\\
    R_{uv}&= R_0/L,\  u\neq v \ .
\end{align}
Same thing for $C_{uv}$:
\begin{align}
    C_{uu}^\infty&= C_d^\infty\\
    C_{uv}^\infty&= C_0^\infty,\  u\neq v \ .
\end{align}
We separate $\hat{\xi}_u$ in a patch independent and a patch dependent part: $\hat{\xi}_u=z\sqrt{C_0^\infty}+w_u \sqrt{C_d^\infty-C_0^\infty}$. 
We call patch disorder average the average over $w_u$ and $\hat{\psi}_u$; species disorder average the average over $z$.
$\frac{1}{L}\sum_v N_v$ concentrates around its average over thermal fluctuations and patch disorder $N^*$, that will be a function of the static Gaussian field $z$.
Substituting in the dynamical equation and using time translational invariance we obtain:
\begin{align}
\label{eq:DMFTfinalapp}
\begin{split}
    \dot{N}=N\left(k-N-\mu h-\sigma\left(\rho \sqrt{C_0^\infty} z+\rho \sqrt{C_d^\infty-C_0^\infty} w+\sqrt{1-\rho^2}\sqrt{C_d^\infty}\hat{\psi}+ \rho\xi + \sqrt{1-\rho^2} \psi\right)\right)+\\+ N\sigma^2\gamma\left( \rho ^2\int_0^t  R_d(t-s)N(s)ds+\rho ^2\int_0^t  R_0(t-s)N^*(s)ds+(1-\rho ^2)\int_0^t  R_d(t-s)N(s)ds \right) +\\+ D (N^*-N) + \eta(t)\sqrt{N}=\\
    =N\left(k-N-\mu h-\sigma\left(\rho \sqrt{C_0^\infty} z+ \sqrt{C_d^\infty-\rho^2 C_0^\infty} w+ \xi \right)\right)+\\+ N\sigma^2\gamma\left( \int_0^t  R_d(t-s)N(s)ds+\rho ^2  R_0^{int}N^* \right) + D (N^*-N) + \eta(t)\sqrt{N} \ ,
\end{split}
\end{align}
where we have summed the random variables that had the same behaviour of the correlations ($w$ and $\hat{\psi}$, $\xi$ and $\psi$),  and
\begin{align}
\begin{split}
    R_0^{int}=\int_0^\infty d\tau R_0(\tau) \ .
\end{split}
\end{align}


The equations simplify in the extreme cases $\rho=1$ and $\rho=0$, because only one of the components of the static part of the disorder is present, either $w$ or $z$. 
For $\rho=1$ $C_d^\infty=C_0^\infty$.
For $\rho=0$ $N^*$ coincides with $h$, so that we have one less self-consistent equation, and $R_0$ is not present; these two facts greatly simplify the numerical solution of the equations.

\section{Stationary probability distribution in the symmetric case}
\label{app:Hamiltonian}

In the case of symmetric interactions ($\gamma=1$), in the single equilibrium phase, the system relaxes to equilibrium and it verifies the Fluctuation-Dissipation Theorem (FDT) \cite{altieri2020}:
\begin{align}
    R_d(\tau)=-\frac{1}{T}\frac{d C(\tau)}{d\tau} \ .
\end{align}
We can integrate by parts the term with the memory kernel:
\begin{align}
    \int_0^t R_d(t-s)N(s)ds = \frac{1}{T} \int_0^t\frac{d C_d(t-s)}{ds} N(s)ds=\\
    \frac{1}{T} \left( C_d^0 N(t)-C(t)N(0)-\int_0^t C_d(t-s) \dot{N}(s)ds \right)=\\
    \frac{1}{T} \left( \left(C_d^0-C_d^\infty\right)N(t) -\int_0^t \left(C_d(t-s)-C_d^\infty\right) \dot{N}(s)ds \right) \ .
\end{align}
We have obtained an additional quadratic term in $N(t)$, and a friction term.
The friction term and the noise $\xi$ describe the coupling of the system to an effective colored bath at temperature $T$, that replaces the coupling of one species to all the others. 

Using Martin-Siggia-Rose-De Dominicis-Janssen (MSRDJ) formalism, we can show that the stationary probability distribution associated with the stochastic differential equation
\begin{align}
\label{eq:DMFTfinalappfriction}
\begin{split}
    \dot{N}=N\left(k-D(1-\rho^2\sigma^2 R_0^{int} N^*)-\mu h-\rho \sigma \sqrt{C_0^\infty} z- \sigma\sqrt{C_d^\infty-\rho^2 C_0^\infty} w-\sigma \xi(t)\right)+\\-N^2\left(1-\frac{\sigma^2}{T}\left(C_d^0-C_d^\infty\right)\right)+ N\frac{\sigma^2}{T}  \int_0^t\left(C_d(t-s)-C_d^\infty\right) \dot{N}(s)ds  + D N^* + \eta(t)\sqrt{N}
\end{split}
\end{align}
is the Boltzmann distribution with the effective Hamiltonian:
\begin{align}
\begin{split}
	H_{eff} = \left( 1-\frac{\sigma^2}{T}\left(C_d^0-C_d^\infty\right) \right)N^2/2+\\ -\left(k- D\left(1-\rho^2\sigma^2 N^* R_0^{int}\right)-\mu h   
 -\rho\sigma \sqrt{C_0^\infty} z- \sigma \sqrt{C_d^\infty-\rho^2 C_0^\infty} w+\zeta\right)N + (T-D N^*+\lambda)\ln N     \ ,
\end{split}
\end{align}
where we have reintroduced the perturbations $\zeta$ and $\lambda$. 
To show that this is the correct equilibrium distribution we need to verify that, with this as an initial condition, time reversal is a symmetry of the associated MSRDJ action. 
We will do it, following reference \cite{aron2010}, for a simplified dynamics, that contains all the crucial ingredients:
 \begin{align}
    \dot{N}=N\left(1-N-\sigma \xi(t) - \sigma^2\int_0^t\nu(t,s)\dot{N}(s)ds\right)  + \eta(t)\sqrt{N}+ \lambda\\
    \langle \xi(t)\xi(s)\rangle= T \nu(t-s)\\
    \langle \eta(t)\eta(s)\rangle= 2 T \delta(t-s) \ .
\end{align}
Its equilibrium distribution is given by:
\begin{align}
    P_{eq}(N)=\frac{e^{-{\beta H}}}{Z}\\
    H=N^2/2-N+(T-\lambda)\log N \ ,
\end{align}
where $\beta=1/T$, the inverse temperature.
The white noise should be interpreted according to Ito's discretization.
It is convenient to convert it to Stratonovich's discretization, which is left invariant by time reversal.
The multiplicative nature of the noise makes the two discretizations not equivalent: we then need to introduce an additional drift term as follows
\begin{align}
    \eta \sqrt{N}\to \eta \sqrt{N} - \frac{1}{2}\frac{\sqrt{2T}}{2\sqrt{N}}\sqrt{2TN}=\eta \sqrt{N} - \frac{T}{2} \ .
\end{align}

The MSRDJ action can be written in terms of a deterministic and a dissipative part \cite{aron2010, aron2016}
\begin{align}
    S[N, \hat{N}]= S^{det}[N, \hat{N}]+ S^{diss}[N, \hat{N}]\\
     S^{det}[N, \hat{N}]=\log P_{eq}(N(-T)) + \int_{-T}^T du\left( i \hat{N}( N(1-N) +\lambda-T/2-T/2)+N-1/2\right)\\
     S^{diss}[N, \hat{N}]=\int_u i \hat{N}_u\int_v (\delta(u-v)+\nu(u-v)\theta(u-v) N_u)(iT\hat{N}_v N_v-\dot{N}_v) \ .
\end{align}

The time reversal transformation for the two fields is given by:
\begin{align}
    N(t)\longrightarrow N_R(t)= N(-t)\\
    i\hat{N}(t)\longrightarrow i\hat{N}_R(t) =i \hat{N}(-t)+\frac{1}{TN(-t)}\frac{\partial}{\partial t}N(-t) \ .
\end{align}

The deterministic and dissipative part of the action are independently invariant under this transformation:
\begin{align}
\begin{split}
 S^{det}[N_R, \hat{N}_R]=-\log Z -\beta H(N(T)) +\\+ \int_u\left(\left( i \hat{N}_{-u}+\frac{1}{TN_{-u}}\frac{\partial}{\partial u}N_{-u} \right)(N_{-u}(1-N_{-u})+\lambda-T)+N_{-u}-1/2\right)=\\
 =-\log Z - \frac{1}{T}(N_T^2/2-N_T+(T-\lambda)\ln N_T)+\\+ \frac{1}{T} \int_u \frac{\partial}{\partial u}\left(N_{u}^2/2 -N_{u}+(T-\lambda)\ln N_u\right) + \int_u\left( i \hat{N}_{u}(N_{u}(1-N_{u})+\lambda-T)+N_u-1/2\right)=\\
 =-\log Z -\beta H(N(-T))  + \int_u\left( i \hat{N}_{u}(N_{u}(1-N_{u})+\lambda-T)+N_u-1/2\right)=S^{det}[N, \hat{N}]
 \end{split}
 \end{align}
\begin{align}
\begin{split}
  S^{diss}[N_R, \hat{N}_R]=\int_u \left( i \hat{N}_{-u}+\frac{1}{TN_{-u}}\frac{\partial}{\partial u}N_{-u} \right)\int_v (\delta_{u-v}+\nu_{u-v}\theta_{u-v} N_u)iT\hat{N}_{-v} N_{-v}=\\
  =\int_u \left( i T \hat{N}_{u}N_u-\dot{N}_{u} \right)\int_v (\delta_{v-u}+\nu_{v-u}\theta_{v-u} N_v)i\hat{N}_{v} = S^{diss}[N, \hat{N}] \ .
 \end{split}
\end{align}
The action is invariant under the time reversal transformation using $P_{eq}$ as initial and final condition, therefore $P_{eq}$ is the correct equilibrium probability distribution. 

\section{Response functions}
\label{app:responsefunction}

In the following, we restrict ourselves to the $\rho=1$ case for simplicity, unless specified, and we show how to obtain the self-consistent equation leading to $R_0^{int}$.

At equilibrium we can rewrite the integrated disorder-dependent responses to a perturbation of the carrying capacity and of the immigration rate in terms of connected correlation functions of $N$. 

\begin{align}
   r_d^{int}(z)=\int_{0}^\infty d \tau\langle\frac{\delta N_u(\tau)}{\delta \zeta_u(0)}\rangle=\frac{\partial \langle N_u\rangle}{\partial \zeta_u}=\beta (\langle N^2\rangle-\langle N\rangle^2)\\
    \chi(z)=\int_0^\infty d\tau \langle\frac{\delta N_u(\tau)}{\delta \lambda_u(0)}\rangle =\frac{\partial \langle N_u\rangle}{\partial \lambda_u}=\beta (\langle N \log N\rangle-\langle N\rangle\langle \log N\rangle) \ .
\end{align}
When the time dependence is not present we are considering a time independent perturbation.

Adding a perturbation in site $v$ leads to a variation of the abundances in all other sites, because of the coupling by diffusion and the memory term. These variations are of order $1/L$, but since there are $L$ of them they give a significant contribution.
When studying $\frac{\partial \langle N_u\rangle }{\partial \zeta_v}$ we need to take into account four contributions: there is a $O(1)$ variation of $N_v$ that leads to a $O(1/L)$ perturbation of the immigration rate perceived by $N_u$ and a $O(1/L)$ change in its off-diagonal memory term; there are $L-2$ variations of $O(1/L)$ of the $N_w$, with $w\neq u, v$, each leading to a $O(1/L^2)$ change in both immigration and memory term.
Carefully taking into account all these contributions, we can write $r_0^{int}(z)$ in terms of $r_d^{int}(z)$, $\chi(z)$ and $r_0^{int}(z)$ itself:
\begin{align}
\begin{split}
    r_0^{int}(z)=L \int_{0}^\infty d \tau\langle\frac{\delta N_u(\tau)}{\delta \zeta_v(0)}\rangle = L \frac{\partial \langle N_u\rangle }{\partial \zeta_v}\\= L  \langle\left(\frac{\partial N_v}{\partial \zeta_v}\left(\frac{D}{L} \frac{\partial N_u}{\partial \lambda_u}+\sigma^2 R_{uv}^{int}\frac{\partial N_u}{\partial \zeta_u}\right)
    + \sum_{w\neq u, v}  \frac{\partial N_w}{\partial \zeta_v}\left(\frac{D}{L} \frac{\partial N_u}{\partial \lambda_u}+\sigma^2 R_{uw}^{int}\frac{\partial N_u}{\partial \zeta_u}\right)\right) \rangle=\\
     =\left(D\chi(z)+ \sigma^2r_d^{int}(z)R_0^{int}\right)\left(r_d^{int}(z)+r_0^{int}(z)\right) \ .
\end{split}
\end{align}
In the third line we used the fact that the correlations between different patches are subleading to take separately the thermal averages. Solving for $ r_0^{int}(z)$ we obtain:
\begin{align}
    r_0^{int}(z)=\frac{\left(D\chi(z)+ \sigma^2r_d^{int}(z)R_0^{int}\right)r_d^{int}(z)}{1-\left(D\chi(z)+ \sigma^2r_d^{int}(z)R_0^{int}\right)} \ .
\end{align}

We can then average over $z$ to obtain $R_0^{int}$:
\begin{align}
     R_0^{int}=\overline{\frac{\left(D\chi(z)+ \sigma^2r_d^{int}(z)R_0^{int}\right)r_d^{int}(z)}{1-\left(D\chi(z)+ \sigma^2r_d^{int}(z)R_0^{int}\right)}} \ .
\end{align}

\section{Asymmetric interactions}
\label{app:asymmetry}

The MSRDJ action with non symmetrical interactions is given by:
\begin{align}
    S[N, \hat{N}]=\int_u i \left(\hat{N}_u\left( N_u\left(k-D\left(1-\sigma^2\gamma  N^*R_0^{int}\right)-\mu h+\sigma \sqrt{C_d^\infty} z-N_u \right)- T+D N^*\right)+N-\frac{1}{2}\right)+\\+\int_u i \hat{N}_u(iT\hat{N}_u N_u-\dot{N}_u) +\frac{\sigma^2}{2}\int_u i \hat{N}_u N_u\int_v C^c(u-v)i\hat{N}_v N_v+\gamma \sigma^2\int_u i \hat{N}_u N_u\int_v R(u-v)N_v +( \log P(N(0)))  ,
\end{align}
where we have defined $C^c(u-v)=C_d(u-v)-C_d^\infty$.
If the introduction of a small asymmetry in the interactions ($\epsilon=1-\gamma\ll 1$) is a non-singular perturbation, all the self-consistently determined quantities in the action ($h$, $C_d$, $R_0^{int}$ and $R_d$) will be close to their counterparts for $\gamma=1$.
At first order in $\epsilon$ we can neglect their change; therefore $R_d$ and $C_d$ will still respect a Fluctuation-Dissipation Relation. 
We can separate the action in a part that would respect FDT and a part that breaks it explicitly:
\begin{align}
    \delta S=  \frac{\epsilon\sigma^2}{T}  \int_{u>v}C^c_{u-v} i \hat{N}_u N_u \dot{N}_v \ .
\end{align}

An average $\langle f(N_t)\rangle $ can be expanded as:
\begin{align}
    \langle f(N_t)\rangle = \langle f(N_t)\rangle_0+\langle f(N_t)\delta S\rangle_0+O(\epsilon^2) \ ,
\end{align}
where $\langle\cdot\rangle_0$ indicates the average with respect to the action neglecting $\delta S$.

We want to estimate the scaling of 
\begin{align}
\label{eq:appdeltaf}
    \langle f(N_t)\delta S\rangle_0= \frac{\epsilon\sigma^2}{T}  \int_{u>v}C^c_{u-v} i \langle f(N_t) \hat{N}_u N_u \dot{N}_v\rangle_0 = \frac{\epsilon\sigma^2}{T}  \int_{u>v}C^c_{u-v} i \frac{\partial }{\partial v}\frac{\delta }{\delta \zeta_u}\langle f(N_t)  N_v\rangle_0
\end{align}
to show that it is not singular approaching a phase transition.
In the simple equilibrium phase the connected correlation function decays exponentially, with a typical relaxation time $\tau$ that could diverge at the phase transitions:
\begin{align}
    C^c(u-v)\sim (\overline{\langle N^2\rangle-\langle N\rangle^2})e^{-|u-v|/\tau} \ .
\end{align}
The correlation function $\langle f(N_t)  N_v\rangle_0$ will contain a $v$ independent part (that we can neglect since we will be taking the derivative in $v$) and a connected component of order 1 that decays with the same relaxation time $\tau$.
Perturbing the system with a field $\zeta_u$ this observable will respond as: 
\begin{align}
    \frac{\delta }{\delta \zeta_u}\langle f(N_t)  N_v\rangle_0\propto \frac{1}{T\tau} e^{-(t-v)/\tau} \ .
\end{align}
Inserting these scalings in equation \ref{eq:appdeltaf} we obtain:
\begin{align}
    \langle f(N_t)\delta S\rangle_0\propto \frac{\epsilon \sigma^2}{T}  \int_{u>v} e^{-(u-v)/\tau}  \frac{\partial}{\partial v}\left(\frac{1}{T \tau} e^{-(t-v)/\tau}\right)
    =  \frac{\epsilon \sigma^2}{T^2\tau^2} \int_{-\infty}^t dv e^{-(t-v)/\tau} \int_{v}^t du e^{-(u-v)/\tau} =\\
    =  \frac{\epsilon \sigma^2}{T^2\tau}  \int_{-\infty}^t dv  e^{-(t-v)/\tau}\left(1-e^{-(t-v)/\tau}\right) = \frac{\epsilon \sigma^2}{T^2\tau}  \left(\tau -\frac{\tau}{2}\right)=\frac{\epsilon \sigma^2}{2T^2}  \ .
\end{align}

Considering a small asymmetry in the interactions observables are shifted by a correction of order $\epsilon$, where the prefactor is of order 1 and has no divergence at the phase transitions.
We thus expected the phase diagram to remain qualitatively unchanged.

\section{Extinction threshold and diversity (for $\rho=1$)}
\label{app:zstar}

\begin{figure}[htbp]
\centering
\includegraphics[width=0.45\textwidth]{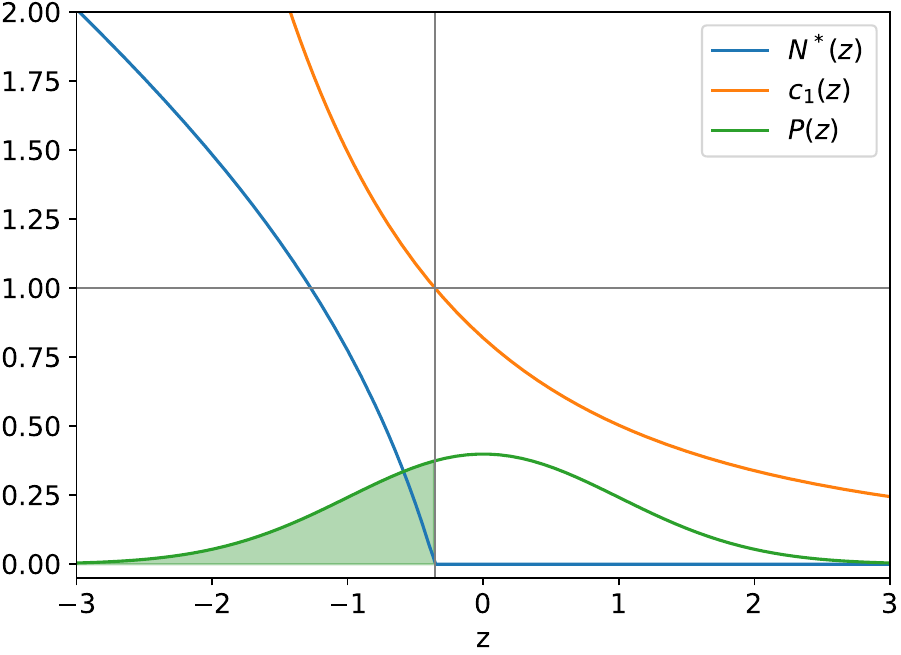}
	\caption{Self-consistent solution for $N^*(z)$ (blue), coefficient of the first order expansion $c_1(z)$ (orange) and Gaussian probability distribution $P(z)$ (green). The highlighted region corresponds to the non-extinct species, its area is the diversity of the ecosystem. $T=0.4$, $D=0.15$, $\mu=1$, $\sigma=0.5$.}
 \label{fig:Nstar}
\end{figure}

The self-consistency condition for $N^*$ reads:
\begin{align}
\label{eq:selfconsNsapp}
    N^*(z)=\langle N \rangle_{H_{eff}(N; h, C_d^0, C_d^\infty, R_0^{int}, z, N^*)}= \frac{\int_0^\infty dN N e^{-\beta H_{eff}(N; h, C_d^0, C_d^\infty, R_0^{int}, z, N^*)}}{\int_0^\infty dN e^{-\beta H_{eff}(N; h, C_d^0, C_d^\infty, R_0^{int}, z, N^*)}} \ .
\end{align}
$N^*=0$ is always a solution of this equation, we want to find the value of $z$ at which it becomes unstable. 

We can separate the effective Hamiltonian into a quadratic and a logarithmic part:
\begin{align}
    H_{eff}(N; h, C_d^0, C_d^\infty, R_0^{int}, z, N^*)= H_q(N; h, C_d^0, C_d^\infty, R_0^{int}, z, N^*)+(T-DN^*)\ln N \ .
\end{align}
For $N^*\to 0$ the logarithmic part gives rise to a non-integrable divergence in 0 in the denominator. 
To improve the numerical stability of our solution at small $N^*$ we performed an integration by parts of the denominator:
\begin{align}
    \int_0^\infty dN e^{-\beta H_{eff}}=\int_0^\infty dN e^{-\beta H_q}N^{-1+\beta DN^*}
    =\frac{1}{ DN^*} \int_0^\infty dN e^{-\beta H_q}N^{\beta DN^*}\frac{d H_q}{dN} \ .
\end{align}
The integral is now finite for $N^*\to 0$ and we can expand eq. (\ref{eq:selfconsNsapp}) in powers of $N^*$:
\begin{align}
\label{eq:selfconsNsappexp}
\begin{split}
    N^*(z)= D N^*(z) \frac{\int_0^\infty dN e^{-\beta H_q}N^{\beta D N^*(z)}}{\int_0^\infty dN e^{-\beta H_q}\frac{d H_q}{dN}N^{\beta D N^*(z)}}
    =N^*(z)  D \frac{\int_0^\infty dN e^{-\beta H_q}}{\int_0^\infty dN e^{-\beta H_q}\frac{d H_q}{dN}}+O((N^*(z))^2) \ .
\end{split}
\end{align}
The term of order $N^*(z)^2$ is always negative, therefore the number of solution depends on the coefficient of the $N^*(z)$ term: if $c_1(z)<1$ the only solution is $N^*(z)=0$, if $c_1(z)>1$ the $N^*(z)=0$ solution is unstable and there is a positive stable one. 
We define the effective growth rate $r_{eff}=1-\sigma^2\beta(C_d^0-C_d^\infty))$ and the effective growth factor $r_{eff} g_{eff}(z)=k-\mu h-z\sqrt{C_d^\infty}\sigma+ D\sigma^2 N^* R_0^{int}$. 
The extinction threshold $z^*$ (Figure \ref{fig:Nstar}) is given by:
\begin{align}
\label{eq:condz*}
\begin{split}
1=c_1(z^*)
 = D\sqrt{\frac{\beta \pi} {2r_{eff}}} \exp\left(\frac{\beta} {2} \frac{(g_{eff}(z^*)r_{eff}-D)^2}{r_{eff}}\right) \left(1+\erf\left(\sqrt{\frac{\beta}{2}}\left(\frac{g_{eff}(z^*)r_{eff}-D}{\sqrt{r_{eff}}}\right)\right)\right) \ .
\end{split}
\end{align}

As noted in reference \cite{denk2022}, this is the same condition that would determine the criticality of the Directed Percolation process with corresponding growth rate and growth factor:
\begin{align}
    \dot{N_u}= r_{eff}(N_u^2/2- g_{eff} N_u) + D\left(\frac{1}{L} \sum_v N_v -N_u\right) + \eta \sqrt{N_u} \\
    \langle \eta(t)\eta(t')\rangle= 2 T\delta(t-t') \ .
\end{align}

The diversity, given by the fraction of non extinct species, can be obtained as:
\begin{align}
    \phi=\int_{z^*}^\infty P(z)dz =\frac{1}{2}\erfc\left(\frac{z^*}{\sqrt{2}}\right) \ .
\end{align}

\section{Continuous transition point}
\label{app:dc}

At the continuous transition, all moments of $N$ tend to zero, and we can expand the extinction condition (\ref{eq:condz*}) in powers of these moments.
Keeping only the zeroth order we obtain an equation on the critical value of the diffusion constant:
\begin{align}
\label{eq:condDc}
D_0\sqrt{\frac{\beta \pi} {2}} e^{\frac{\beta} {2} (k-D_0)^2} \left(1+\erf\left(\sqrt{\frac{\beta}{2}}\left(k - D_0\right)\right)\right)=1 \ .
\end{align}
This condition has no dependence on the distribution of the interactions, indeed it is the same that would be obtained with zero or constant interactions \cite{denk2022}.

For $T\to 0$ we can expand Eq. (\mbox{\ref{eq:condDc}}) and show that $D_0$ vanishes exponentially:
\begin{align}
    D_0(T)\sim \frac{1}{\sqrt{2 \pi\beta}} e^{-\frac{k^2}{2T}}
\end{align}
The reason for this behavior is that at low demographic noise, the abundances of a species with carrying capacity $k$ undergoes a fluctuation toward very low values very rarely. In fact, one needs to wait a rare fluctuation of the demographic noise that makes the species go against the force due to the logistic growth. This phenomenon is similar to the one encountered in the Kramers' problem for barrier crossing. Using the same line of arguments employed there, one finds that the timescale for this rare event is $e^{\frac{k^2}{2T}}$ (the "energy barrier" equals $k^2/2$). The equation above can be therefore interpreted as a balance between two inverse time-scales: the one needed for diffusion to operate and the one over which extinctions take place.

By a careful (and cumbersome) expansion of the self-consistent equations, we can show that approaching the continuous transition $h\propto q_d\propto D-D_0$, $ q_0\propto (D-D_0)^2$, and $z^*$ (and therefore $\phi$) has a finite limit.

\section{Abundance distribution}
\label{app:PN}

\begin{figure}[htbp]
\centering
\begin{subfigure}[b]{0.3\textwidth}
\hfill
         \centering
         \includegraphics[width=\textwidth]{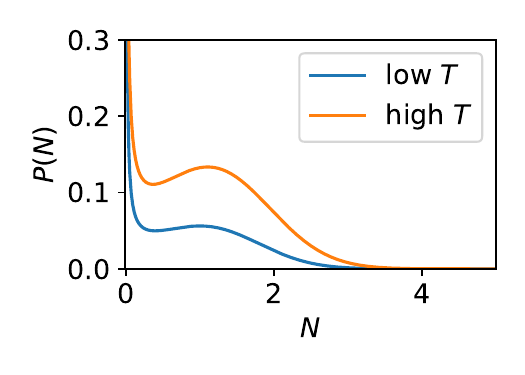}
         \caption{}
         \label{fig:abdistr}
     \end{subfigure}
     \hfill
     \begin{subfigure}[b]{0.3\textwidth}
         \centering
         \includegraphics[width=\textwidth]{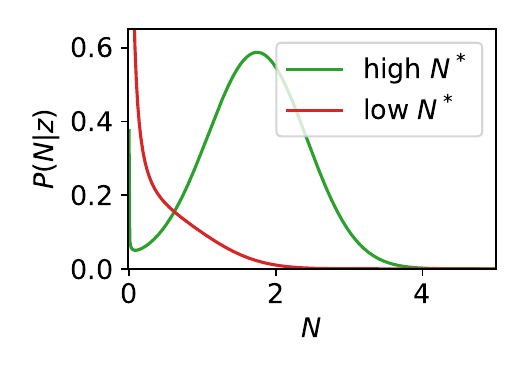}
         \caption{}
         \label{fig:abundancedistributiongivenz}
     \end{subfigure}
     \hfill
     \begin{subfigure}[b]{0.3\textwidth}
         \centering
         \includegraphics[width=\textwidth]{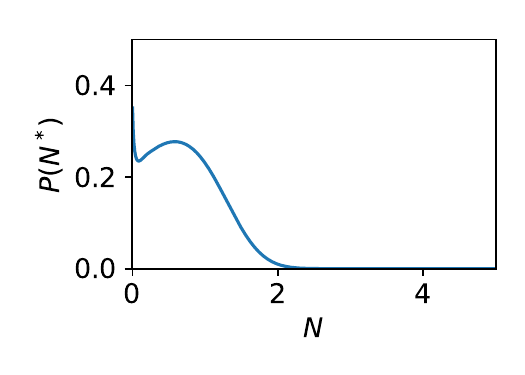}
         \caption{}
         \label{fig:averageabundancedistribution}
     \end{subfigure}
     \hfill
\caption{(a) Probability distribution of the abundance $N$; in orange deep in the survival phase ($T=0.8, D/D_0(T)=1.5$), in blue right before the discontinuous transition ($T=0.4$, $D/D_0(T)=0.84$). Note that the shown distributions do not integrate to 1 because a finite fraction of the species are extinct, leading to a delta function in zero with weight $1-\phi$. (b) Probability distribution of the abundance for a given species, i.e. at fixed $z$; in blue for a species close to extinction ($z=-0.45$, $N^*=0.148$), in orange for a species far from extinction ($z=-2.45$, $N^*=1.733$); $T=0.4$, $D/D_0(T)=0.84$. (c) Probability distribution of the space (or time) averaged abundance, because of extinct species we again have a delta function in zero with weight $1-\phi$. $T=0.4$, $D/D_0(T)=0.84$. $\mu=1$, $\sigma=0.5$, $\rho=1$.}
	\label{fig:abundancedistribution}
\end{figure}

As noted before, two types of stochasticity contribute to the distribution of abundances. 
Each species is subjected to demographic and environmental noise, making their abundance a time-dependent random variable. 
For each species the abundance is distributed according to the Boltzmann distribution with Hamiltonian $H_{eff}$, we will call this $P(N|z)$. 
On top of this, because of disorder, different species experience different average interactions with the rest of the ecosystem (different values of $z$, that is distributed according to $P(z)$ Gaussian), leading to species-dependent factors in $H_{eff}$. 
If we want to study the distribution of the abundances of all species at a given time in one site ($P(N)$) we need to take into account both effects.
We can compute $P(N)$ marginalizing over $z$:
\begin{align}
    P(N)=\int dz P(N|z) P(z)  \ ,
\end{align}
with 
\begin{align}
    P(N|z)=\frac{ e^{-\beta H_{eff}(N; z, N^*(z))}}{\int_0^\infty dN e^{-\beta H_{eff}(N; z, N^*(z))}}\\
    P(z)=\frac{e^{-z^2/2}}{\sqrt{2 \pi}} \ .
\end{align}

We could also be interested in the distribution across species of the abundance averaged over patches or time, given by $P(N^*)=P(z)\left(\frac{dN^*(z)}{dz}\right)^{-1}$ for $N^*>0$. There is also a finite probability $1-\phi$ that $N^*=0$, where $\phi$ is the diversity. 

Examples of these abundance probability distributions are shown in Fig. \ref{fig:abundancedistribution}.

\section{Divergence of response to a variation of the carrying capacity}
\label{app:precursor}

The divergence of response functions when approaching a tipping point is a generic feature of saddle node bifurcations \cite{scheffer2012anticipating}.
Let us consider a generic dynamical system, described by 
\begin{align}
    \frac{d \vec{x}}{dt}=F(\vec{x}, k) \ .
\end{align}
$\vec{x}$ contains all the degrees of freedom of the system, whereas $k$ is a control parameter. 
The zeros of $F$ yield the stationary states $x^*$:
\begin{align}
    F(x^*, k)=0 \ .
\end{align}
The stationary point is stable if the Jacobian of $F$ has only negative eigenvalues, ensuring that $x$ returns to $x^*$ upon small perturbations. 
In a saddle node bifurcation a stable and an unstable stationary point collide and annihilate each other. 
Since the Jacobian at the stable stationary point has only negative eigenvalues whereas the one at the unstable stationary point has at least a positive eigenvalue, one of the eigenvalues has to cross zero at the bifurcation.
The existence of a zero mode leads to a diverging response to perturbation.

We show below how this mechanism is at play in our case when approaching the stability limit of the self-sustained phase. 
We study the response of the system to a perturbation in the environmental conditions in the case of independent interaction coefficients ($\rho=0$). 
We will consider for concreteness a perturbation to the carrying capacity $k$, but we expect the same qualitative behavior for perturbations to the diffusion constant, the moments of the interactions, or the strength of the demographic fluctuations. 

The response of the order parameters to a variation of $k$ involves some connected moments of $N$ and the derivative of $H$ in $k$:
\begin{align}
\label{eq:dhdk}
\begin{split}
    \frac{dh}{dk}=\int \mathcal{D}z\left(\frac{\int dN Ne^{-\beta H}(-\beta)dH/dk}{\int dN e^{-\beta H}}-\frac{\int dN Ne^{-\beta H}\int dN e^{-\beta H}(-\beta)dH/dk}{\left(\int dN e^{-\beta H}\right)^2}\right)=\\
    =-\beta\left(\overline{\langle N\frac{dH}{dk}\rangle}-\overline{\langle N\rangle \langle\frac{dH}{dk}\rangle}\right)
\end{split}
\end{align}
\begin{align}
    \frac{dC_d^0}{dk}=-\beta\left(\overline{\langle N^2\frac{dH}{dk}\rangle}-\overline{\langle N^2\rangle \langle\frac{dH}{dk}\rangle}\right)\\
\label{eq:dq0dk}
    \frac{dC_d^\infty}{dk}=-2\beta\left(\overline{\langle N\rangle\langle N\frac{dH}{dk}\rangle}-\overline{\langle N\rangle^2 \langle\frac{dH}{dk}\rangle}\right) \ .
\end{align}

Thanks to the fact that $\rho=0$, $H$ has the simplified form:
\begin{align}
	H= \left( 1-\sigma^2 \beta\left(C_d^0-C_d^\infty\right) \right)N^2/2 +(\mu h-k+D-z \sqrt{C_d^\infty}\sigma)N + (T-D h)\ln N \ .
 \end{align}
 
$dH/dk$ depends on the derivative of the order parameters in $k$:
\begin{align}
	\frac{dH}{dk}=-N+\frac{\partial H}{\partial h}\frac{d h}{dk}+\frac{\partial H}{\partial C_d^0}\frac{d C_d^0}{dk}+\frac{\partial H}{\partial C_d^\infty}\frac{d C_d^\infty}{dk}
 \end{align}
\begin{align}
    \frac{\partial H}{\partial h}&=\mu N-D\ln N\\
    \frac{\partial H}{\partial C_d^0}&=-\frac{\sigma^2\beta}{2}N^2\\
    \frac{\partial H}{\partial C_d^\infty}&=\frac{\sigma^2\beta}{2}N^2-\frac{1}{2\sqrt{C_d^\infty}}z\sigma N \ .
\end{align}

Substituting $dH/dk$ in Equations (\ref{eq:dhdk}-\ref{eq:dq0dk}) we obtain:
\begin{align}
\label{eq:dhdk2}
\begin{split}
    \frac{dh}{dk}=-\beta\Big\{-\left(\overline{\langle N^2\rangle}-\overline{\langle N\rangle^2 }\right)+\\+\left[\mu \left(\overline{\langle N^2\rangle}-\overline{\langle N\rangle^2 }\right)-D\left(\overline{\langle N\log N \rangle}-\overline{\langle N\rangle\langle \log N\rangle}\right)\right]\frac{dh}{dk}+\\
    -\frac{\sigma^2\beta}{2}\left(\overline{\langle N^3\rangle}-\overline{\langle N\rangle\langle N^2\rangle }\right)\frac{dC_d^0}{dk}+\\
    +\left[\frac{\sigma^2\beta}{2}\left(\overline{\langle N^3\rangle}-\overline{\langle N\rangle\langle N^2\rangle }\right)-\frac{1}{2\sqrt{C_d^\infty}}\sigma \left(\overline{\langle N^2\rangle z}-\overline{\langle N\rangle^2 z }\right) \right] \frac{dC_d^\infty}{dk}
    \Big\}
\end{split}
\end{align}
\begin{align}
\begin{split}
    \frac{dC_d^0}{dk}=-\beta\Big\{-\left(\overline{\langle N^3\rangle}-\overline{\langle N^2 \rangle \langle N\rangle }\right)+\\
    +\left[\mu \left(\overline{\langle N^3\rangle}-\overline{\langle N^2 \rangle \langle N\rangle }\right)-D\left(\overline{\langle N^2\log N \rangle}-\overline{\langle N^2\rangle\langle \log N\rangle}\right)\right]\frac{dh}{dk}+\\
    -\frac{\sigma^2\beta}{2}\left(\overline{\langle N^4\rangle}-\overline{\langle N^2\rangle^2 }\right)\frac{dC_d^0}{dk}+\\
    +\left[\frac{\sigma^2\beta}{2}\left(\overline{\langle N^4\rangle}-\overline{\langle N^2\rangle^2 }\right)-\frac{1}{2\sqrt{C_d^\infty}}\sigma \left(\overline{\langle N^3\rangle z}-\overline{\langle N^2\rangle\langle N\rangle z }\right) \right] \frac{dC_d^\infty}{dk}
    \Big\}
\end{split}
\end{align}
\begin{align}
\label{eq:dq0dk2}
\begin{split}
    \frac{d C_d^\infty}{dk}=-2\beta\Bigg\{-\left(\overline{\langle N\rangle\langle N^2\rangle}-\overline{\langle N\rangle^3 }\right)+\\+\left[\mu \left(\overline{\langle N\rangle\langle N^2\rangle}-\overline{\langle N\rangle^3 }\right)-D\left(\overline{\langle N\rangle\langle N\log N \rangle}-\overline{\langle N\rangle^2\langle \log N\rangle}\right)\right]\frac{dh}{dk}+\\
    -\frac{\sigma^2\beta}{2}\left(\overline{\langle N\rangle\langle N^3\rangle}-\overline{\langle N\rangle^2 \langle N^2\rangle }\right)\frac{dC_d^0}{dk}+\\
    +\left[\frac{\sigma^2\beta}{2}\left(\overline{\langle N\rangle \langle N^3\rangle}-\overline{\langle N\rangle^2\langle N^2\rangle }\right)-\frac{1}{2\sqrt{C_d^\infty}}\sigma \left(\overline{\langle N\rangle\langle N^2\rangle z}-\overline{\langle N\rangle^3 z }\right) \right] \frac{dC_d^\infty}{dk}\Bigg\} \ .
\end{split}
\end{align}
We collect the three order parameters in a vector $\vec{p}=(h, C_d^0, q_0)^T$. Then $\frac{d\vec{p}}{dk}$ satisfies:
\begin{align}
    \frac{d\vec{p}}{dk}=\hat{J}\frac{d\vec{p}}{dk}+ \vec{s} \ ,
\end{align}
where $\hat{J}$ is a $3\times 3$ matrix and $s$ a vector; their elements are the coefficients of Equations (\ref{eq:dhdk2}-\ref{eq:dq0dk2}).
The solution is given by:
\begin{align}
    \frac{d\vec{p}}{dk}=-(\hat{J}-\hat{1})^{-1} \vec{s} \ .
\end{align}
The response to a variation of $k$ diverges if $\hat{J}-\hat{1}$ has a zero eigenvalue, this is found to happen when approaching the discontinuous transition.

We expect the same qualitative behaviour of the response to perturbations for generic values of $\rho$, but for $\rho\neq 0$ we need to take into account also the variations of the function $N^*(z)$, which leads to the study of an infinite dimensional matrix.

\section{Reduced interaction matrix}
\label{app:restrictedmatrix}

\begin{figure}[htbp]
\centering
\begin{subfigure}[b]{0.24\textwidth}
\hfill
         \centering
         \includegraphics[width=\textwidth]{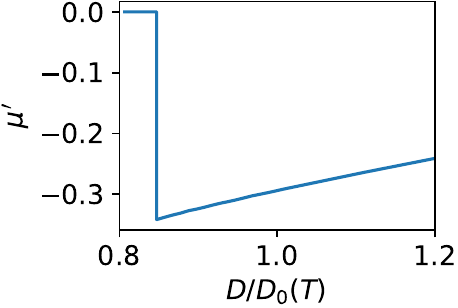}
         \caption{}
         \label{fig:muprime}
     \end{subfigure}
     \begin{subfigure}[b]{0.25\textwidth}
         \centering
         \includegraphics[width=\textwidth]{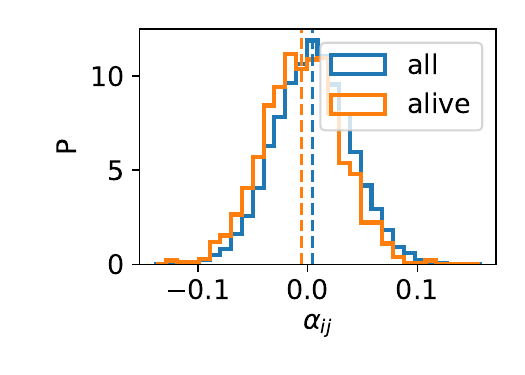}
         \caption{}
         \label{fig:alphanum}
     \end{subfigure}
     \hfill
\caption{Left: Analytical estimate of the mean of the reduced interaction matrix using Eq. (\ref{eq:muprime}) for $T=0.4$. Right: Numerical results for the distribution of the interaction coefficients for all and surviving species. $T=0.18$, $D/D_0=0.8$.  In both cases in the initial species pool $\mu=1$, $\sigma=0.5$; if all species go extinct we say $\mu'=0$.}
	\label{fig:reducedalpha}
\end{figure}

In the case of fixed interaction matrices, a finite fraction of the species goes extinct; the interaction matrix restricted to the surviving species has a smaller mean than the starting one.
The statistics of the reduced interaction matrix can be computed at 0 temperature \cite{baron2023}:
\begin{align}
\label{eq:muprime}
    \mu'=\phi \mu -2\frac{\sigma^2 h}{\phi} \frac{d \phi}{d \zeta} =\phi \mu -2\frac{\sigma h}{\phi \sqrt{2 \pi C_d^\infty}} e^{-(z^*)^2/2}  \ .
\end{align}
Since we are not at 0 temperature, in our case this formula is only an approximation, but it provides an useful estimate of the variation of the mean interaction.
We find that the interaction mean decreases (more mutualistic) when decreasing the diffusion coefficient (Figure \ref{fig:muprime}); it is negative in the entire metastability region.   
In Figure \ref{fig:alphanum} we show the distribution of the interaction coefficients considering all species or only surviving ones in numerical simulations.
The distribution of the interaction coefficients is slightly shifted to more negative values, and indeed $\mu'=S \frac{1}{S^2} \sum_{ij}\alpha_{ij}$ changes from 0.96 to -0.28.

To compute the average interaction term we can again use the cavity method and imagine to add a species (with index 0) to the community. Using Equation \ref{eq:intDMFT}
\begin{align}
    Int_0 = \langle \sum_j \alpha_{0j} N_j^u\rangle= \mu h  +\sigma \sqrt{C_d^\infty}z - \gamma \sigma^2\left(R_d^{int} +R_0^{int}\right) \langle N_0\rangle \ .
\end{align}
We can now average it over all species (all values of $z$, overline), or over only non extinct ones ($z<z^*$, overline with $^+$ superscript).
\begin{align}
    \overline{I}=\mu h- \gamma \sigma^2\left(R_d^{int} +R_0^{int}\right) h\\
    \overline{I}^+=\mu h -\sigma \frac{\sqrt{q_0}}{\phi}\frac{e^{-z^{*2}/2}}{\sqrt{2 \pi}} - \gamma \sigma^2\left(R_d^{int} +R_0^{int}\right) \frac{h}{\phi} \ .
\end{align}
Note that we will always find $\overline{I}^+<\overline{I}$; $\overline{I}^+$ is negative in the entire metastability region (Figure \ref{fig:intAn} in the main text).
This is also confirmed by numerical simulations: the average interaction term is 0.13 considering all species, and -0.46 considering only non extinct ones (Figure \ref{fig:intNum} in the main text).

In the case of independent interaction matrices, all species survive, so that the interaction matrix is not modified.

\section{Numerical scheme}
\label{app:numerical}

The numerical simulation of demographic noise poses some technical challenges. 
Naively sampling it as a Gaussian variable can result in negative species abundances, an unphysical result that makes the scheme numerically unstable.
A clever solution was found in reference \cite{dornic2005}, and improved in \cite{weissmann2018, altieri2021}. 
The idea is to separate the process in a deterministic part:
\begin{align}
    \dot{N}_{i,u}=N_{i,u}\left(1-N_{i,u}-\sum_j\alpha_{ij}^u N_{j,u} \right)+D\left(\frac{1}{L}\sum_v N_{i,v}-N_{i,u}\right)         
\end{align}
and a stochastic one:
\begin{align}
    \dot{N}_{i,u}=\sqrt{N_{i,u}}\eta_{i,u} \ .
\end{align}
At each time step we numerically integrate the two in sequence. 
For the stochastic part an exact solution of the associated Fokker-Planck equation is available for any initial condition, and it can be efficiently sampled using Gamma and Poisson variables:
\begin{align}
    \tilde{N}_{i,u}(t)=Gamma\left(Poisson\left(\frac{N_{i,u}(t)}{T dt}\right)\right)T dt \ .
\end{align}
For the deterministic part we rely on Euler method.
\begin{align}
    N_{i,u}(t+dt)=\left(\tilde{N}_{i,u}(t)\left(1-\tilde{N}_{i,u}(t)-\sum_j\alpha_{ij}^u \tilde{N}_{j,u}(t) \right)+D\left(\frac{1}{L}\sum_v \tilde{N}_{i,v}(t)-\tilde{N}_{i,u}(t)\right) \right) dt       \ .
\end{align}

\section{Additional numerical results}
\label{app:numres}

Some of the challenges encountered in numerical simulations become clear examining the time evolution of the average abundances (Fig. \ref{fig:hvst}). At high temperature (top) the average abundance fluctuates significantly even with large number of species and patches ($S=200$, $L=400$); finite size effects on $L$ lead to an excess of extinctions. At lower temperature (bottom) the dynamics strongly slows down, and at $t=200$ some of the abundances (depending on the value fo $D$) have not yet reached their asymptotic value, leading to a smoothing of the discontinuous transition.

\begin{figure}[htbp]
\centering
\begin{subfigure}[b]{0.49\textwidth}
         \centering
         \includegraphics[width=\textwidth]{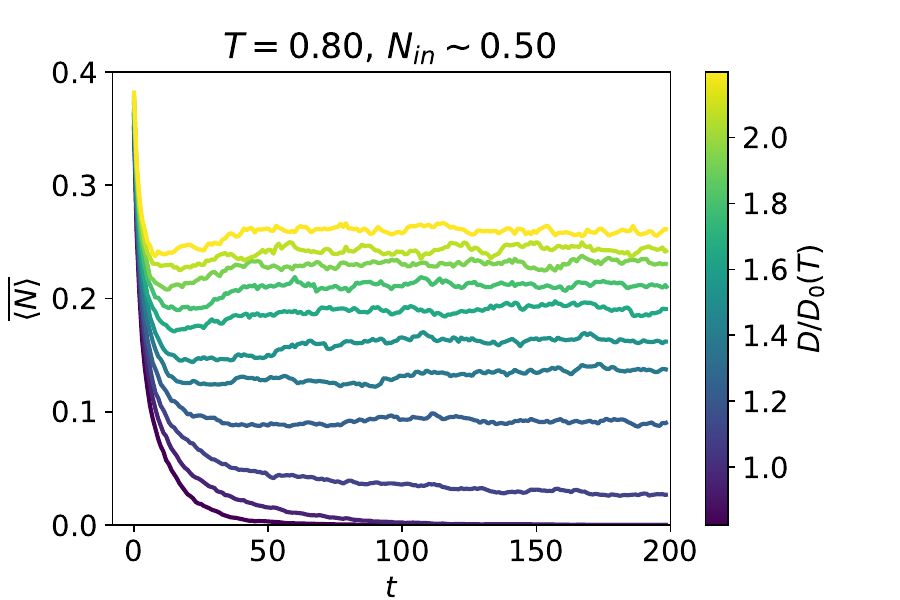}
         \label{fig:hT041}
     \end{subfigure}
     \hfill
     \begin{subfigure}[b]{0.49\textwidth}
         \centering
         \includegraphics[width=\textwidth]{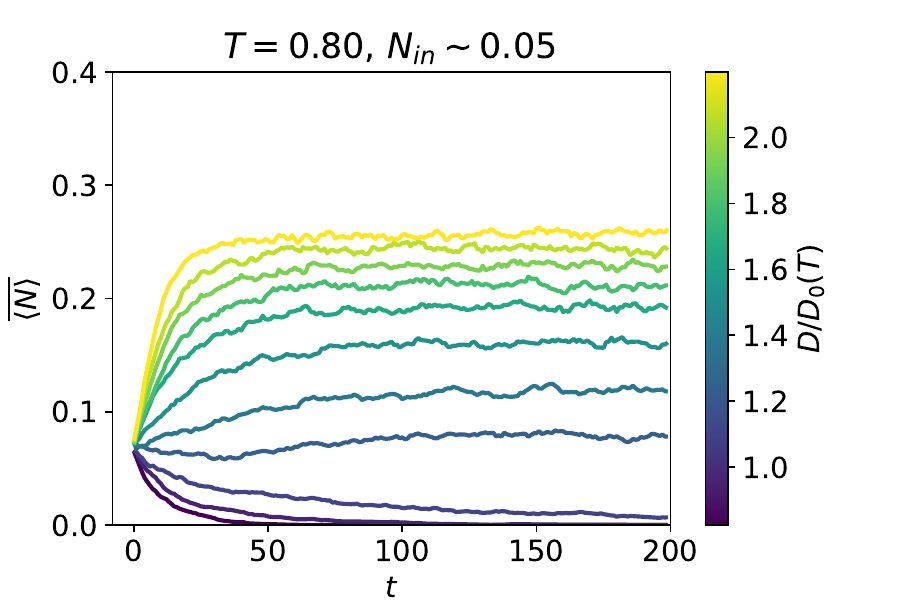}
         \label{fig:hT021}
     \end{subfigure}
     \hfill
\begin{subfigure}[b]{0.49\textwidth}
         \centering
         \includegraphics[width=\textwidth]{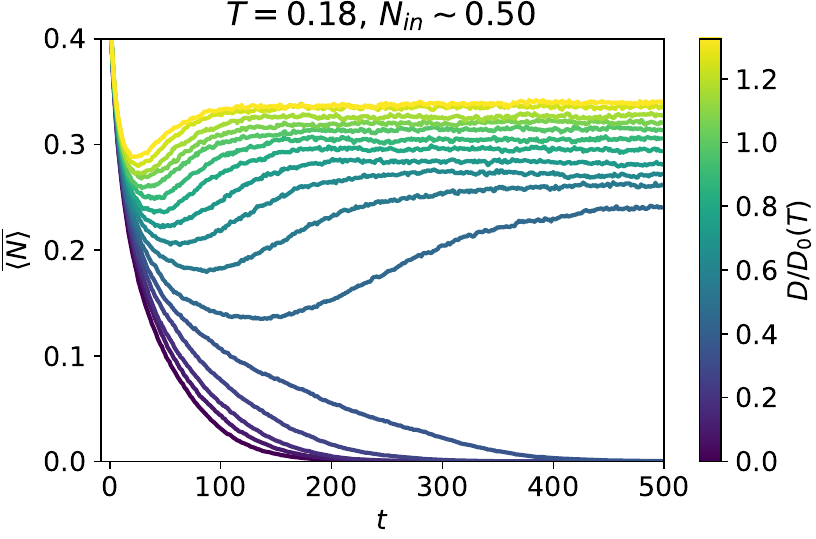}
         \label{fig:hT042}
     \end{subfigure}
     \hfill
     \begin{subfigure}[b]{0.49\textwidth}
         \centering
         \includegraphics[width=\textwidth]{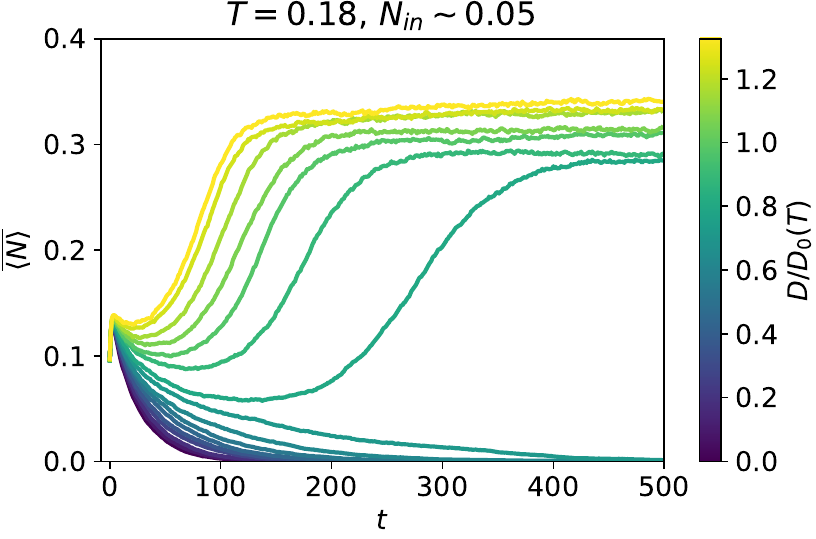}
         \label{fig:hT022}
     \end{subfigure}
     \hfill
\caption{Time evolution of the average abundance for two different temperatures and two average values of the initial conditions. Note the different time ranges in the top and bottom figures: at high temperature the abundances have converged to their asymptotic values at $t_{max}=200$, at lower temperature it is necessary to wait much longer ($t_{max}=500$). $S=200$, $L=400$, $\mu=1$, $\sigma=0.5$.}
	\label{fig:hvst}
\end{figure}

\begin{figure}[htbp]
\centering
     \begin{subfigure}[b]{0.49\textwidth}
         \centering
         \includegraphics[width=\textwidth]{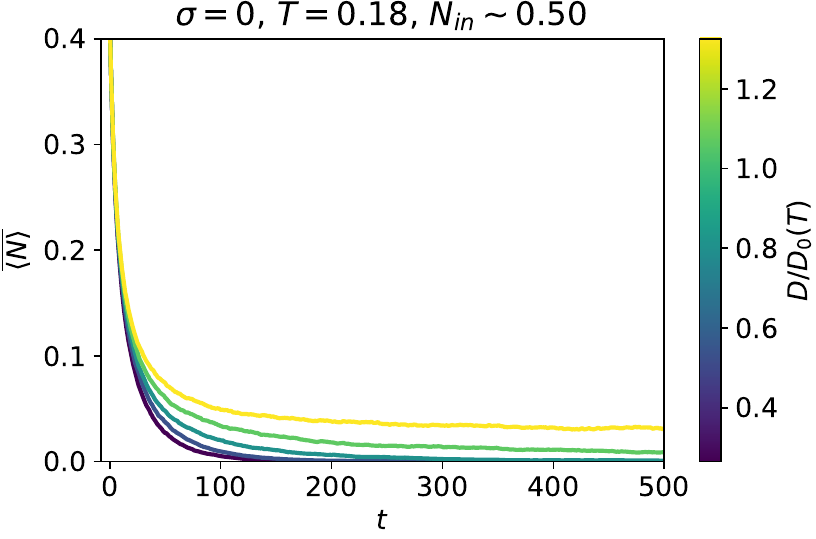}
         \label{fig:hsigma01}
     \end{subfigure}
     \hfill
     \begin{subfigure}[b]{0.49\textwidth}
         \centering
         \includegraphics[width=\textwidth]{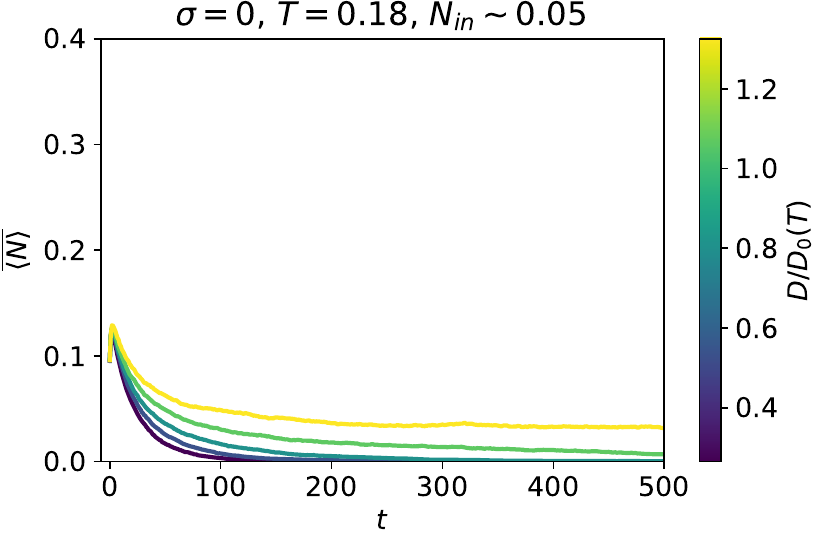}
         \label{fig:hsigma022}
     \end{subfigure}
     \hfill
\caption{Time evolution of the average abundance without heterogeneity in the interaction network ($\sigma=0$) at $T=0.18$ and two average values of the initial conditions. For $D<D_0(T)$ the abundances converge to 0.  $S=200$, $L=400$, $\mu=1$. }
	\label{fig:hvstsigma0}
\end{figure}

\begin{figure}[htbp]
\centering
     \begin{subfigure}[b]{0.49\textwidth}
         \centering
         \includegraphics[width=\textwidth]{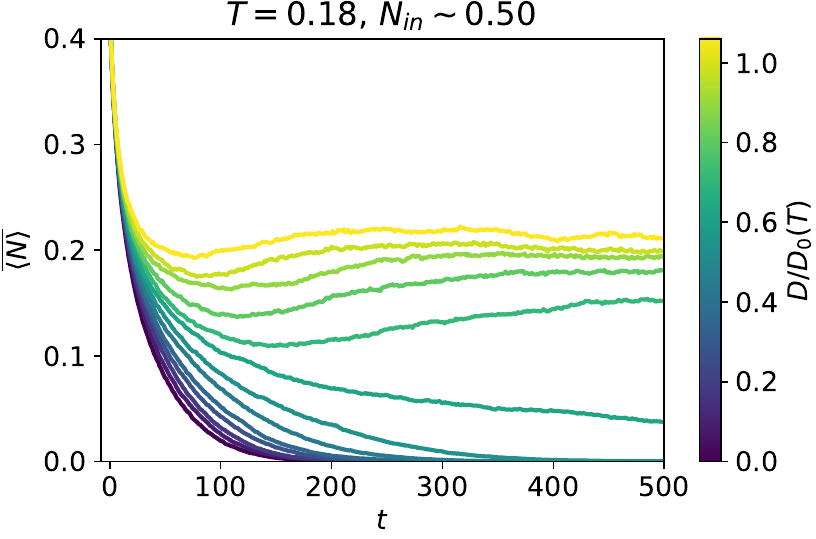}
         \label{fig:hgamma1}
     \end{subfigure}
     \hfill
     \begin{subfigure}[b]{0.49\textwidth}
         \centering
         \includegraphics[width=\textwidth]{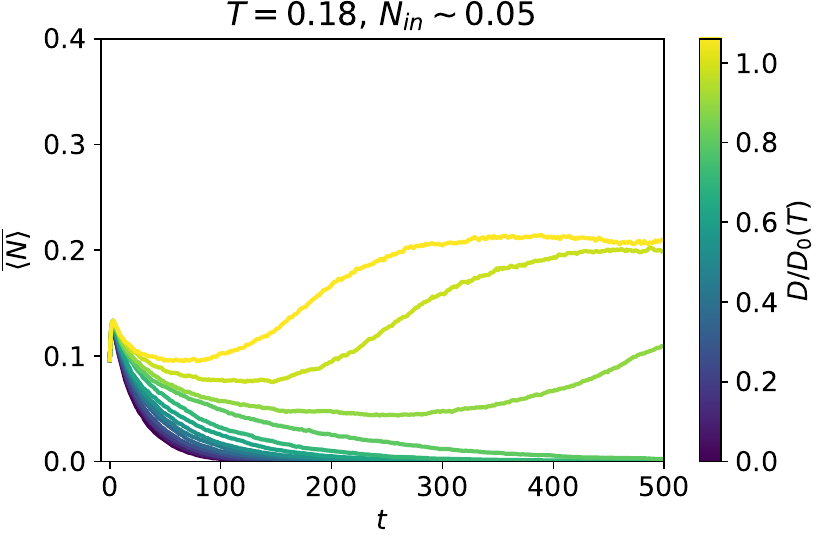}
         \label{fig:hgamma2}
     \end{subfigure}
     \hfill
\caption{Time evolution of the average abundance without heterogeneity in the interaction network with partial correlation between patches ($\rho=0.9$) and non symmetric interactions ($\gamma=0.9)$ at $T=0.18$ and two average values of the initial conditions. At $t=500$ the abundances have not yet reached their asymptotic value, leading to an apparent smoothing of the discontinuous transition. Nevertheless this is ensured by the abrupt change of behaviour of the evolution of the average abundance: for one value of the diffusion constant at long times the abundance is decaying to 0, whereas for the next it shows a (slow) increase. We conclude that the asymptotic values would likewise show an abrupt change. $S=200$, $L=400$, $\mu=1$, $\sigma=0.5$.}
	\label{fig:hvstgamma}
\end{figure}

\newpage
\end{widetext}
\end{document}